\documentclass[a4paper,useAMS,usenatbib]{mn2e}

\newcommand{\de}{$^{\circ}$}

\newcommand{\mf}{magnetic field}

\newcommand{\mfs}{magnetic fields}

\newcommand{\pangle}{$| \chi_\mathrm{B} |$}

\newcommand{\etaeq}{$\eta=\,$}
\newcommand{\dd}{\mathrm{d}}
\newcommand{\expp}{\mathrm{exp}}

\newcommand{\mapsCaptionA}{
Left: polarisation maps. Vectors follow $\chi_\mathrm{B}$ and their length
is given by $p$. Vectors are superimposed on linear contours of
$I/\left< I \right>$. Right: logarithmic grayscale maps of $I/\left< I \right>$.}

\usepackage{graphicx}
\usepackage{subfigure}

\title[3D-MHD simulations of evolving magnetic fields in FR~II radio sources]{
Three-dimensional magnetohydrodynamic simulations of the evolution of
magnetic fields in Fanaroff-Riley class~II radio sources
}

\author[M. Huarte-Espinosa, M. Krause and P. Alexander]{M. 
Huarte-Espinosa,$^{1,2,5}$\thanks{E-mail: martinhe@pas.rochester.edu} 
M. Krause,$^{3,4}$ and P. Alexander,$^{2,5}$ \\
$^1$Department of Physics and Astronomy, University of Rochester, 600 Wilson Boulevard, Rochester, NY, 14627-0171 \\
$^2$Astrophysics Group, Cavendish Laboratory, 19~J.~J.~Thomson~Ave., Cambridge CB3 0HE, UK \\
$^3$Max-Planck-Institut f\"ur Extraterrestrische Physik, 
Giessenbachstrasse, 85748 Garching, Germany \\
$^4$Universit\"atssternwarte M\"unchen, Scheinerstr.~1, 81679 M\"unchen, Germany \\
$^5$Kavli Institute for Cosmology Cambridge, Madingley Road, Cambridge CB3 0HA, UK}
\begin{document}

\date{Received \today}
\pagerange{\pageref{firstpage}--\pageref{lastpage}} \pubyear{2009}
\maketitle
\label{firstpage}

\begin{abstract}
Radio observations of Fanaroff-Riley class~II sources often show
correlations between the synchrotron emission and the linear-polarimetric
distributions.  Magnetic position vectors seem to align with the
projected emission of both the radio jets and the sources' edges.
Using statistics we study such relation as well as its unknown time
evolution via synthetic polarisation maps of model FR~II sources
formed in 3D-MHD numerical simulations of bipolar, hypersonic and
weakly magnetised jets. The magnetic field is initially random with
a Kolmogorov power spectrum, everywhere.  We investigate the structure
and evolution of magnetic fields in the sources as a function of
the power of jets and the observational viewing angle. Our synthetic
polarisation maps agree with observations, showing B-field vectors
which are predominantly aligned with the jet axis, and show that
magnetic fields inside sources are shaped by the jets' backflow.
Polarimetry is found to correlate with time, the viewing angle and
the jet-to-ambient density contrast. The magnetic structure inside
thin elongated sources is more uniform than inside more spherical
ones. We see jets increase the magnetic energy in cocoons in
proportion to the jet velocity 
and the cocoon width. 
Filaments in the synthetic emission maps suggest
turbulence develops in evolved sources.
\end{abstract}


\section{Introduction}
\label{intro}

Centimetric wavelength observations reveal synchrotron emission from
extragalactic Fanaroff-Riley Class~II radio sources (\citealp{FR};
FR~IIs hereafter) and radio-loud quasars (see e.g. \citealp{bridle84}
and references therein).  Linear polarisation fractions \hbox{within
$\sim$\,10--50\,\%} are commonly seen in these objects. Polarisation
maps of these sources show patchy distributions which correlate with 
the luminosity distribution (see \citealp{saikia88} for a review).  
Projected magnetic field
vectors are predominantly parallel to both the radio jets and the 
boundaries of 
radio lobes (Alexander, Brown \& Scott 1984;
\citealp{bridle84}; Leahy, Pooley \& Riley 1986,~1997; \citealp{black92};
Johnson, Leahy \& Garrington 1995; \citealp{hardcastle97},~1998;
\citealp{leahy97,gilbert04,mullin06}). Strong emission gradients
are often followed by the vectors perpendicularly, and when multiple
hotspots are observed in one of the two~radio lobes, the vectors seem
to follow a line that would connect the hotspots. Linear polarisation 
fractions of radio jets tend to be higher at the edges than at inner 
regions.
Linear polarisation fractions of radio lobes tend to be higher at the edges 
than at regions both inside and between the lobes \citep{saikia88}.

The direction of the \mf\ component that is in the plane of the sky is
often inferred by computing the Stokes parameters on the observed signal.
It is possible to do the calculations inversely in order to model the
polarimetry distribution that results from given magnetic field geometries
\citep{laing81a,jones88}. Such studies indicate
that magnetic fields in FR~IIs seem to  
%
   consist of 
	a combination of ordered and disordered (anisotropic) 
	fields along the jets and their vicinities, as well as 
   a random component at the inner regions of radio lobes
   \citep{laing81b,bridle84,saikia88}.
%
Circumferential magnetic structures are also frequently observed in the
outer edge of the sources \citep{bridle84,saikia88}.

Based on the radio luminosity distribution observed in several
FR~IIs, \citet{rees71}, Longair, Ryle \& Scheuer (1973),
\citet{blandford74} and \citet{scheuer74} proposed the following
model for the plasma dynamics in the sources.  Magnetised relativistic
plasma jets are launched from a central engine located inside active galactic
nuclei (AGN) which are typically seen at positions that match
those of the radio cores. 
 %
 %
A cavity (the
cocoon, hereafter) is inflated with the jets' plasma and a strong
bow~shock is driven on the ICM. Jets collide with the ambient medium
at their working surfaces. Radio hotspots are seen 
at the leading edges of the lobes
because the plasma pressure is the highest there. 
The plasma nearby is pushed
towards the radio core and a backflow of magnetised plasma develops. 
Radio synchrotron lobes are thus formed, separated from the 
ICM by a contact discontinuity.

At kiloparsec-scales, \mfs\ in FR~IIs are
often modeled in energy equipartition with the synchrotron emitting
ultra-relativistic electrons. 
Magnetic flux freezing is expected to 
bond field lines and radio emitting electrons dynamically, hence 
the jets' backflow should play an important role in shaping magnetic fields
inside cocoons \citep{laing80,alexander84,leahy84,miller85,saikia88}.

The expansion of radio sources must be considered to understand
their inferred magnetic structure.  Evolutionary models have provided
analytical expressions for the global time dependence of the volume,
the pressure and the energy inside 
cocoons (\citealp{scheuer74,
falle91,begelman98,kaiser97,vlj1}; Heinz, Reynolds \& Begelman  1998).
The large-scale features of the complex non-linear dynamics in such plasma
cavities have been captured by numerical simulations (for reviews see
\citealp{norman93,ferrari98,pudritz07}). Two-dimensional axisymmetrical
simulations of magnetised jets have confirmed the basic picture regarding
jets (beams), lobes (cocoons)
and bow shocks (e.g. Clarke, Norman \& Burns 1986; K\"ossl, M\"uller \&
Hillebrandt 1990a,~b; \citealp{lind89,frank98,stone00,komissarov99}).
These simulations also show that cocoons consist of a series of vortices.
These structures arise in a complex feedback loop, where 
pressure modulations in the
cocoons interact with the beams' shock pattern which, in turn,
modifies the vortex shedding. The vortices decay in a turbulent cascade
\citep[compare section~3 in][]{ka07}. The latter process results in
some degree of isotropisation of the field lines which should affect
the alignment of magnetic fields and the fractional
polarisation. 

Further, the expansion of cocoons involves magnetic field amplification.
This happens 
via two field line stretching processes
\citep{matthews90b}: The poloidal stretching mechanism, which arises because
the fluid elements in the beam located close to the beam boundary take 
small turns, 
and 
thus 
end up towards the inner part of 
cocoons. 
In contrast, the 
fluid elements close to the jet axis make larger turns and 
end up near the
outer cocoon boundary. Because of the larger path length of these
fluid elements, they lag behind. Hence
shear amplifies the magnetic
field in cocoons along the direction of the jet axis. 
On the other hand, 
the toroidal stretching process amplifies the toroidal component
of magnetic fields via cocoon expansion perpendicular to the jet
axis. Unless the flow structure is axisymmetric, which is an unlikely
configuration for real radio sources, the toroidal magnetic field
may again be sheared, and thereby converted into poloidal field.
To first order, the resulting magnetic field structure is 
determined by both, 
these competing processes and the initial condition.
%
  This picture has been refined by Gaibler, Krause \& Camenzind (2009)
  who initialised their simulation with a helical magnetic field confined
  to the beam. 
%
The poloidal component of these fields returns to the
source along, and close to, the beam, and therefore 
its strength drops steeply
with distance from the jet axis, $R$. The radial cocoon expansion
puts work into the toroidal field which, consequently, increases linearly
with $R$, as predicted by \citet[][toroidal stretching]{matthews90b}.
Hence, the magnetic energy in radio lobes could be largely due to 
dynamo action in the large scale jet flow, with little
dependence on the conditions (set up) at the base of the beam. 
The literature on three-dimensional jet simulations, in contrast,
has not paid much attention to these issues. In general, one finds 
jet instabilities which are transparent to simulations with less 
degrees of freedom, e.g. jet fluting, deflection, disconnection and 
splash-back (see e.g. \citealp{norman93}).
%
%
%
%
	Jet propagation has been studied also with relativistic MHD
	codes (e.g. \citealp{leismann05,keppens,mignone},
	and references therein). These studies
	have not particularly focussed on polarisation properties.
	Relativistic jets have narrower cocoons for a given rest
	mass density ratio, and more stable beams. The motions in
	radio lobes are subrelativistic, and hence their physics
	should not be too much influenced by a relativistic nature
	of the jet.
%

Synthetic observations are produced using data from numerical
simulations in order to compare them with observations.
%
   \citet{matthews90a} simulated the hydrodynamic advection and polarized
   synchrotron emission of random, pasive \mfs\ in AGN jets, finding high linear
   polarisation fractions, of about 70\,\%.
%
\citet{clarke93} carried out 3D simulations of
the interaction of a jet and a cloud with passive uniform \mfs.
The synthetic synchrotron emission maps of Clarke showed filaments,
formed by velocity shear.  \citet*{hardee97} carried out
3D-MHD simulations of supermagnetosonic magnetised perturbed
equilibrium beams, where a section of an infinitely long beam is 
studied, and found
synthetic intensity structures similar to the ones observed in
the jets of Cygnus A. More recently, Tregillis, Jones \& Ryu 
%
	(2004b)
	investigated the fractional polarisation of 
   synthetic synchrotron observations of 3D-MHD
	   AGN 
	jet simulations.
	They found 
	   rather high 
	fractional polarisations
   in regions where shock acceleration increases the emissivity,
	    but much smaller 
	fractional polarisation 
   at regions where relativistic particles illuminate the volume more
   uniformly.  
%
In general, little attention has been given to the
statistics of synthetic polarimetry and the way it relates with the
properties of radio jets.

In this paper we present 3D-MHD numerical simulations of hypersonic
magnetised jets as well as synthetic
synchrotron and polarisation observations. 
In contrast to Gaibler et al. (2009) and much other work, we do not
start with a regular magnetic field component within the jet, but rely entirely
on the field amplification due to the dynamics of 
cocoons (compare above)
to create structure. Regarding analysis and the questions we address,
we follow essentially \citet{matthews90a} with the important
improvement that here we use a full 
three-dimensional 
magnetohydrodynamic treatment for the jet simulation.
%

This paper is organised as follows: in Section~\ref{simul} we
describe the formalism of ideal MHD and the numerical methods we
use. Our implementation of the ICM, CMFs and AGN jets are also
described there along with details of our calculations for the
synthetic synchrotron emission and polarimetry.  In Section~\ref{results}
we talk about the flow structure in our model sources and analyse
it in terms of energetics. Synthetic maps are then presented and
compared with FR~II radio observations.  The results
are then interpreted and analysed statistically. Section~\ref{discu}
is dedicated to compare our models with previous numerical simulations.  
We then summarise
and conclude our study in Section~\ref{conclu} which is followed by
the bibliography.

\section[]{Simulations}
\label{simul}


We describe the dynamics of plasma in the ICM and AGN radio jets
using the system of nonlinear time-dependent hyperbolic equations of 
ideal compressible MHD. In three~dimensions and 
non-dimensional conservative form, these are given by:

\begin{eqnarray} 
   \label{eq:mass}
   \frac{\partial\rho}{\partial t} + \nabla\cdot(\rho\mbox{\bf V}) &= 
   \dot{\rho}_\mathrm{j}~~~~~~~~&
   \\
   \label{eq:momentum}
   \frac{\partial (\rho {\bf V})}{\partial t} + 
   \nabla\cdot \left( \rho {\bf V V} + p_g + B^2/2 - 
   {\bf B B}\right)  &= \rho {\bf g} + \dot{\bf P}_\mathrm{j}&
   \\
   \label{eq:energy}
   \frac{\partial E}{\partial t} + \nabla\cdot\left[\left(E
   +p_g + B^2/2 \right)
   {\bf V}-{\bf B}({\bf V}\cdot{\bf B})\right] &= 
   \dot{E}_j~~~~~~~& 
   \\
   \label{eq:induction}
   \frac{\partial {\bf B}}{\partial t} - \nabla\times( {\bf V}
   \times {\bf B}) &= 0,~~~~~~~& 
\end{eqnarray}
\noindent where $\rho$, $p_g$, {\bf V} and {\bf B} are the plasma density,
thermal pressure, flow velocity and magnetic fields, respectively.
In (\ref{eq:energy}), \hbox{$E=p_g/(\gamma-1)+\rho V^2/2+B^2/2$} and represents
the total energy density whereas $\gamma$ is the 
ratio of specific heats.  In the right hand side of (\ref{eq:mass}),
(\ref{eq:momentum})~and~(\ref{eq:energy}) source terms are used to
implement jets by injecting mass, $\dot{\rho}_j$, momentum,
$\dot{P}_j$, and kinetic energy, $\dot{E}_j$ (see Section~\ref{jets}),
as well as a Newtonian gravitational acceleration, ${\bf g}$, to
keep the plasma in magneto-hydrostatic equilibrium (see Section~\ref{icm}).

We solve the above equations 
in three~dimensions
using the numerical code Flash~3.1 \citep{fryxell00}.
Flash's new
multidimensional unsplit constrained transport solver is employed to 
maintain the divergence of magnetic fields down 
\hbox{to $\la$\,10$^{-12}$} \citep{lee08}.
A diffusive HLLC solver \citep{hllc} prevents spurious low pressure
and density values from appearing in the grid.
We use a Courant-Friedrichs-Lewy parameter of~0.25 and
periodic boundary conditions in all the domain's faces.
These boundary conditions prevent numerical noise from polluting
the turbulent magnetic spectrum in the grid (Section~\ref{cmfs}).
Our computational domain is a cube with edges 
$|{\bf x}| \le 1/2$, in computational~units, 
   and has a uniform grid with 200$^3$~cells.
   This represents a volume of 200\,kpc$^3$ meant to 
   simulate the core of a cluster.

We carried out five jet simulations (see Table~1)
designed to experiment with the power of jets
in terms of their velocities and densities. 
%
%
%
Computations ran for approximately \hbox{12\,hours} on 64 processors 
at the CamGrid\footnote{http://www.escience.cam.ac.uk/projects/camgrid/} 
cluster of the University of Cambridge, and the production runs executed 
for about \hbox{4\,hours} (using 64 processors) at the 
Darwin\footnote{http://www.hpc.cam.ac.uk/darwin.html} supercomputer of 
the University of Cambridge HPC facility.

\subsection[]{Initial conditions} 
\label{init}

\subsubsection[]{The ICM} 
\label{icm}

The cluster plasma is implemented using an equation of state of an
ideal monoatomic gas, with a ratio of specific heats $\gamma=5/3$,
a constant sound speed (\hbox{$c_s^2 = \gamma \, p_g / \rho = 1$})
throughout the domain and a density following a King profile
\citep{king72}

\begin{equation}
\rho_\mathrm{ICM}(r) = \frac{\rho_\mathrm{c}}{(1+(r/r_\mathrm{c})^2)^{3\beta/2}},
\label{king}
\end{equation}
\noindent where the central density, $\rho_\mathrm{c}$, the central radius, $r_\mathrm{c}$, 
and $\beta$ take the values of 1, 0.8 and \hbox{2$/$3}, respectively.

To keep the magnetised gas in magneto-hydrostatic
equilibrium we implement a radial acceleration source term ${\bf g}$
to equation~(\ref{eq:momentum}), and take the balance between this
term and the total plasma pressure $p_g+B^2/2$. In the radial direction 
this term takes the form:
\begin{equation}  
   g_\mathrm{r} =  -\frac{ 2 \, c_\mathrm{s}^2 }{ \gamma r_\mathrm{c}^2 } \frac{r}{( 1 + (r/r_\mathrm{c})^2) }
   \left( 1 + \frac{1}{\beta_\mathrm{m}} \right),
\label{grav}
\end{equation}  
\noindent where $\beta_\mathrm{m}$ is the ratio of thermal pressure to
magnetic pressure.

\subsubsection[]{Cluster \mf} 
\label{cmfs}

The \mf\ within the cluster is set up as an isotropic random field
with a power law energy spectrum.
Following \citet{tribble91b} and \citet{murgia04} 
we generate a cubic grid in Fourier space,
  with 200$^3$\,cells. For each of these,
we define three components of a vector potential which takes the
form ${\bf \tilde{A}}({\bf k})\,=\,{\bf A}({\bf k})e^{i {\bf \theta}({\bf
k})}$, where ${\bf k}$ is the frequency vector
($k^2=k_x^2+k_y^2+k_z^2$), $i$ is the unitary complex
number, while ${\bf A}$ and ${\bf \theta}$ are the vector amplitudes and
phases, respectively.  We draw ${\bf \theta}({\bf k})$ from a uniform
random distribution within 0 and 2$\pi$, and ${\bf A}({\bf k})$ is also
randomly distributed but has 
a Rayleigh probability distribution
\begin{equation}
   P(A,\theta)\dd A\dd \theta = \frac{A}  {2 \pi |A_k|^2}
   \expp \left( - \frac{A^2}  {2 |A_k|^2} \right) \dd A \dd \theta,
   \label{eq:Rayleigh}
\end{equation}
\noindent where we choose the power law Ansatz
\begin{equation}
   |A_k|^2 \propto k^{- \zeta},
   \label{eq:zeta}
\end{equation}
for a given slope $\zeta$.

We transform to real space by taking the inverse fast Fourier Transform
\citep{recipes} of ${\bf \tilde{A}}({\bf k})$. 
The resulting 
magnetic vector potential, ${\bf A}({\bf x})$,
is multiplied 
by the plasma density
radial profile~(\ref{king}). This product implements magnetic
flux freezing by generating fields, 
the strength of which 
follows the plasma density,
and pressure, profile.
The components of the vector potential are 
then 
read and mapped into
the staggered-grid cell interfaces of Flash3.1 and the curl of
this vector is then calculated to give the magnetic field. 
Finally, we normalise the resulting
field so that the ICM's thermal pressure is approximately
ten~times larger than its magnetic pressure (\hbox{$\beta_m =
p_g/(B^2/2) \, \sim \, 10$}) everywhere in the grid, which is a 
reasonable value in this context
\citep{carilli02}.

This procedure yields solenoidal \mfs\ 
tangled at scales
of order our computational resolution 
and characterized by spatial 
variations following a magnetic power spectrum with a power law of the form
\begin{equation}
   |B|^2 \propto k^{- \zeta + 2} = k^{-n},
   \label{eq:POW}
\end{equation}
\noindent where we choose a Kolmogorov three-dimensional turbulent
slope \hbox{$n= -$11$/$3}, based on the work of 
\citet{vogt03,vogt05b} and \citet{guidetti08}. 
We use the same realisation for all our runs.
%
   We note that the Fourier method implicitly imposes maximum and minimum
   scale on the field.

We let this plasma relax for one crossing time before injecting the jets.

\subsection{Jets}
\label{jets}

By implementing source terms to equations~(\ref{eq:mass}),
(\ref{eq:momentum})~and~(\ref{eq:energy}) we inject mass, 
$x-$
momentum and kinetic energy to the central grid cells that are
within a control cylinder of radius $r_j$ and height $h_j$, 
resolved by 3~and \hbox{8\,cells}, respectively. Inside this ``nozzle'' we
update the plasma density and $x$-velocity 
via 
constant source terms $\dot{\rho}_\mathrm{j}$ and $\dot{v}_\mathrm{j}$.
Jets are continuously injected until they reach the computational
boundaries and then the simulations are stopped.  Plasma pressure in the nozzle, $p_j$,
takes the constant value of the central ambient pressure (i.e.
\hbox{$\rho_c/\gamma$}). The jet density is computed using $\rho_j
= \eta \rho_c$, where the parameter $\eta$ takes the (low) values given
in Table~1. 
%
  We assume an ideal gas equation of state with $\gamma=\,$5$/$3
  for the jet material.
%
The light densities of our jets are motivated by the work of 
\citet{alexander96} and 
\citet{vlj1}, and their high Mach numbers are based on the 
observed jets sidedness associated with Doppler
beaming, suggesting that FR~II sources are 
at least close to relativistic up to scales of order \hbox{100\,kpc}
\citep{mullin09}.
%
   The Mach numbers of our jets with respect to the sound speed in the
	ambient gas are 40, 80 and 130 which
   correspond to velocities close to 66$\times$10$^3$,
   133$\times$10$^3$ and 216$\times$10$^3$\,km\,s$^{-1}$, respectively.
   The Mach numbers of the jets with respect to the sound speed in
	the beam material are 2.5, 5.7, 5, 11.3 and 8.2, as they appear
	in Table~1.
%
We extend the implementation of \citet{omma04} to simulate bipolar
magnetised jets. The launch and collimation of the jets are
assumed to occur in the AGN ``central engine'' located at
sub-resolution scales.

The initial jet magnetic fields are kept from the initialisation of
the ambient medium, and no magnetic source term is applied. 
It therefore has a random topology, an average 
\hbox{$\beta_\mathrm{m} \sim\,$10}
%
   and, given the assumed power spectrum (Section~\ref{cmfs}), it
   is fairly uniform at scales $\sim r_j$.  We note, however, there
   is no reason to believe the magnetic fields in FR~II radio jets 
	are related to the CMFs near the AGN; jet fields are expected to 
	be advected up the beam from the central engine.  Our choice of 
	initial jet magnetic fields is based on the fact they seem to be weak
	at kiloparsec-scales and to have a random component (Section~1).
   This is the case of the central fields in our model.
%

As our jets propagate, their
magnetic fields are deformed by shear. The time averaged average
beam \hbox{$\beta_\mathrm{m}$} is of about~50. 
The power of jets is the sum of thermal and kinetic 
power terms:
\begin{equation}
   L_j = \int \left( \frac{1}{2} ( \eta \rho_\mathrm{c} ) v_j^2 \right)
	   v_x \, dA + \int \left( \frac{\gamma p_j }{\gamma -1} \right) 
	   v_x \, dA,
\label{jetpower0}
\end{equation}
\noindent which takes the following form at the grid
\begin{equation}
   L_j  =  \frac{1}{2} ( \eta \rho_\mathrm{c} ) (\pi r_j^2) v_j^3 +
       \frac{\gamma}{\gamma -1}~p_j (\pi r_j^2) v_j.
   \label{jetpower}
\end{equation}

\setcounter{table}{0}
\begin{table}
\centering
    \begin{minipage}{80mm}
   \caption{Simulations and parameters.}
   \begin{tabular}{@{}lccccc@{}}
   \hline
   Simulation
      &$v_j$\footnote{Time average average jet velocity in the nozzle. 
			It is equal to the external Mach~number.}
         &$\eta$\footnote{Time average average jet to ambient density 
                          contrast in the nozzle.}
	    &$L_j$\footnote{Jet power from equation~(\ref{jetpower}).}
               &$t_e$\footnote{Simulation end times.} \\    
   name  
      &[Mach] 	
          &
	    &[$\times$10$^{38}$\,W]  
               &[Myr]  \\    
   \hline           
   lighter-slow		 &~40    &0.004   &~~\,4.6    &14.1   \\
   light-slow		    &~40    &0.020   &~\,17.2  &~\,8.3   \\
   lighter-fast		 &~80    &0.004   &~28.1    &~\,7.1   \\
   light-fast		    &~80    &0.020   &128.8    &~\,4.4   \\
   lighter-faster     &130    &0.004   &112.8    &~\,4.7   \\
   \hline
   \end{tabular}
   \end{minipage}
\label{table1}
\end{table}
 
\subsubsection{Cocoon contact surface}
\label{cocoon}

We use a passive incompressible tracer, $\tau({\bf
x},t)$, which is injected with the jet plasma to distinguish
it from that of the ambient medium. When jet injection starts,
$\tau({\bf x},t_\mathrm{jet}=\,$0)
takes the values of 0.99 and 1$\times$10$^{-10}$ at the nozzle and 
at the ICM, respectively.  The tracer is then advected with
the jet gas and takes values within 1$\times$10$^{-10}$ to 0.99. A
comparison of the distributions of $\tau$ and $\rho$ allows us
to 
identify the contact surface of the cocoon with an accuracy up to 4
computational cells. 
%
%

  \begin{figure*}
    \centering
       {\bf ~Mach$=\,$80, $\eta=\,$0.004 \& $t_{jet}=\,$7.1\,Myr} 
       ~~~~~~~~~~~~~~~~~~~~~~~~~~~~~~~
       {\bf Mach$=\,$80, $\eta=\,$0.02 \& $t_{jet}=\,$4.4\,Myr} \\ 
   \vskip.2cm
          \includegraphics[width=.9\textwidth]{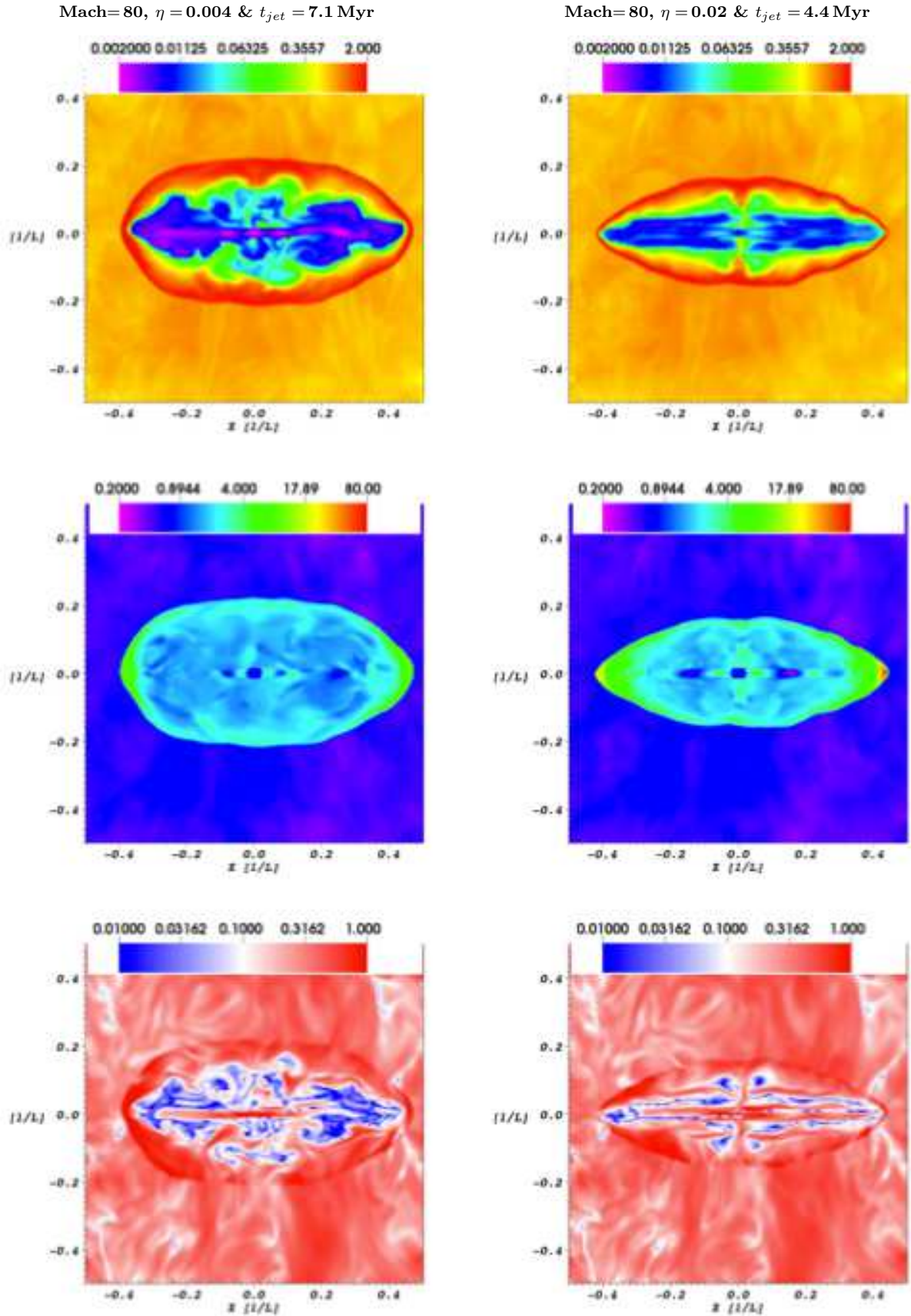}        
   \caption{Two dimensional cuts ($z=\,$0) of the
  density (top row), the pressure (middle row) and the magnetic
  field strength (bottom row) distributions. The Mach number and density
  of the runs are given on the top of each column.  Colour scales
  are logarithmic and show variables in corresponding computational
  units. We see a clear relation between the intrinsic structure
  of sources and the resultant field structure in the synthetic
  polarisation maps
  (Figures~\ref{pol-sim01-90deg-a}--\ref{pol-sim04-90deg-a}).}
       \vspace*{0pt}
  \label{2dcuts}
  \end{figure*}

 \begin{figure*}
    \centering
    \includegraphics[width=1\textwidth]{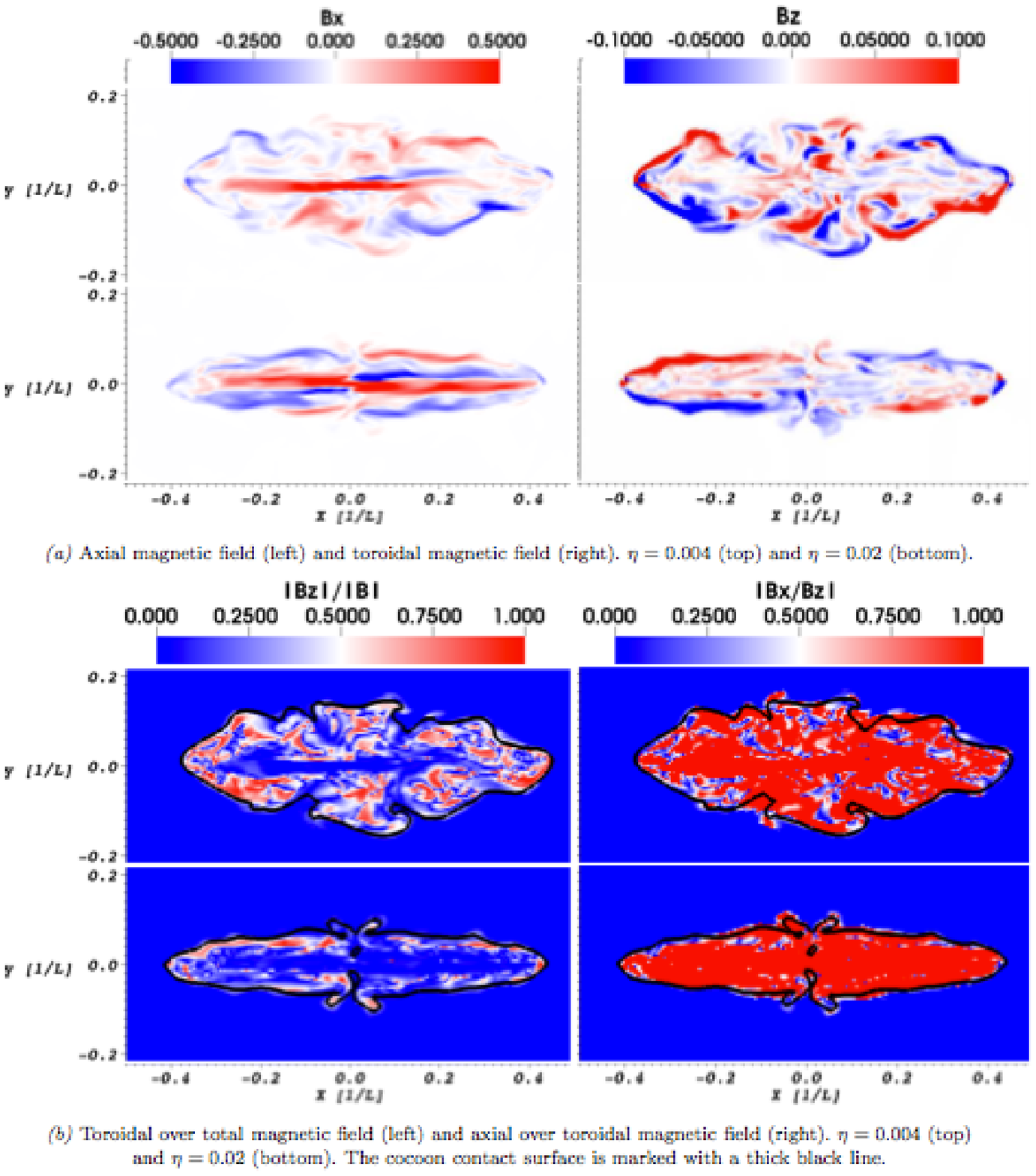}
    \caption{Two dimensional cuts ($z=\,$0) of 
	 the magnetic fields for the Mach~80 jets (see Table~1).
    Colour scales are linear and show variables in corresponding computational units.}
    \vspace*{0pt}
  \label{2dcutsB}
  \end{figure*}

\subsection[]{Synthetic radio maps}
\label{maps}

Our simulations produce three-dimensional data cubes with information
about the distribution of the magnetised gas in 
our 
model sources 
at different times during their expansion. 
Synthetic synchrotron emission and polarimetry
are computed under the assumption
that the radiation is linearly polarized. 
%
%
Beaming and light-travel effects are assumed to be negligible.
Synthetic observations are
produced at viewing angles, $\theta_\mathrm{v}$, of 30, 60 and 90~degrees
measured from the jet axis to the line of sight 
(thus jets are in the plane of the sky when $\theta_\mathrm{v}=\,$90$^{\circ}$).
Given a viewing angle and a simulation timestep,~$t$, Stokes parameters
are (i) calculated for every computational cell inside the
source, 
using the \mf\ components in the plane of the sky,
${\bf B'}({\bf x},t)$, (ii) integrated through the source, along
the line of sight, $Z_{\parallel}(t)$.  Mathematically,
\begin{equation}
  \begin{array}{ll}
  I({\bf x}_{\perp},t) ~~= ~~\frac{1}{l} \int_0^l 
  	&\delta(\tau) \, \tau({\bf x},t) \, p_c({\bf x},t) 
          \, [ B'_x({\bf x},t)^2 +  \\
          &B'_y({\bf x},t)^2 ] \,  
          dZ_{\parallel}(t), 
  \\
  Q({\bf x}_{\perp},t) ~= ~\frac{0.75}{l} \int_0^l 
  	&\delta(\tau) \, \tau({\bf x},t) \, p_c({\bf x},t) 
          \, [ B'_x({\bf x},t)^2 - \\
          &B'_y({\bf x},t)^2 ] \,
          dZ_{\parallel}(t), 
  \\
  U({\bf x}_{\perp},t) ~= \frac{0.75}{l} \int_0^l 
  	&\delta(\tau) \, \tau({\bf x},t) \, p_c({\bf x},t) \, 2\,
             \, B'_x({\bf x},t) \\
             &B'_y({\bf x},t) \, 
             dZ_{\parallel}(t),
  \end{array} 
\label{stokes}
\end{equation}
\noindent where 
\begin{equation}
\delta(\tau) = \left\{
\begin{array}{c l}
    1  & \rmn{for} \,\, \tau({\bf x},t) \in [0.5,.99] 
       \,\,\rmn{(cocoon)}; \\ 
    0  & \rmn{for} \,\, \tau({\bf x},t) \in [1 \times 10^{-10},0.5) 
       \,\,\,\,\,\,\rmn{(ambient)},
\nonumber
\end{array}
    \right.
\nonumber
\end{equation}
  \noindent and ${\bf x}_{\perp}, p_c({\bf x},t)$
  and $I$ represent the coordinates in the plane of the sky, the
  distribution of the 
  cocoon pressure and the total intensity of the radiation,
  respectively.  
  %
     We note (\ref{stokes}) are valid for a synchrotron emission
     spectral index $\alpha=\,$1, yet the degree of
     polarisation we predict does not vary too much with $\alpha$
     \citep{laing80}. 
     The factor of~0.75 
  %
  in the expression of $Q$ and $U$
  in (\ref{stokes}) accounts for the maximum degree of linear
  polarisation for a uniform magnetic field and a power-law electron
  energy distribution. 
  %
	  We model the density distribution of synchrotron emitting
	  electrons via the factor $\tau({\bf x},t) \, p_c({\bf x},t)$ in 
	  (\ref{stokes}). We do not
	  follow any explicit energy gain or loss processes such
	  as synchrotron cooling or shock acceleration  (i.e. the
	  background plasma pressure is proportional to the constant
	  factor in the energy distribution of relativistic electrons).
	  A detailed treatment of the electron distribution is
	  beyond the scope of this paper. We note that we have also
	  tried a constant density of radiating electrons, which
	  did not significantly change the results.
     %
     %
  %
%
 The polarisation angle of the magnetic vectors, $\chi_\mathrm{B}$,
and the degree of linear polarisation ($=$ fractional polarisation), $p$, are given by
\begin{equation}
  \chi_\mathrm{B} = \frac{1}{2} \, \arctan(U/Q) + \frac{\pi}{2}, \qquad 
  p = \frac{\sqrt{U^2+Q^2}}{I}.
  \label{chi}
\end{equation}
\noindent

   \begin{figure}
\includegraphics[width=.475\textwidth] {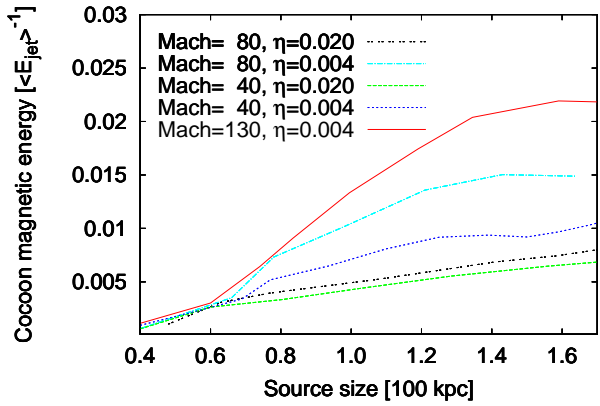}
   \caption{Time evolution of the magnetic energy in the cocoon.
Only the 
$\eta=\,$0.004 
sources show flat gradients, typical of MHD turbulent dynamos.
}
   \label{cocoon-ene}
   \end{figure}


\section[]{Results and analysis}
\label{results}

Synthetic polarisation and emission maps are presented in pairs
characterized by the jet velocity (same as the external Mach number),
the density contrast $\eta$, the viewing angle $\theta_\mathrm{v}$
and the time that jets have been active,
$t_{\rmn{jet}}$. Polarisation maps have a constant vector density
of \hbox{0.5\,cells$^{-2}$}.

\begin{figure*}
	\centering
         \includegraphics[width=\textwidth]{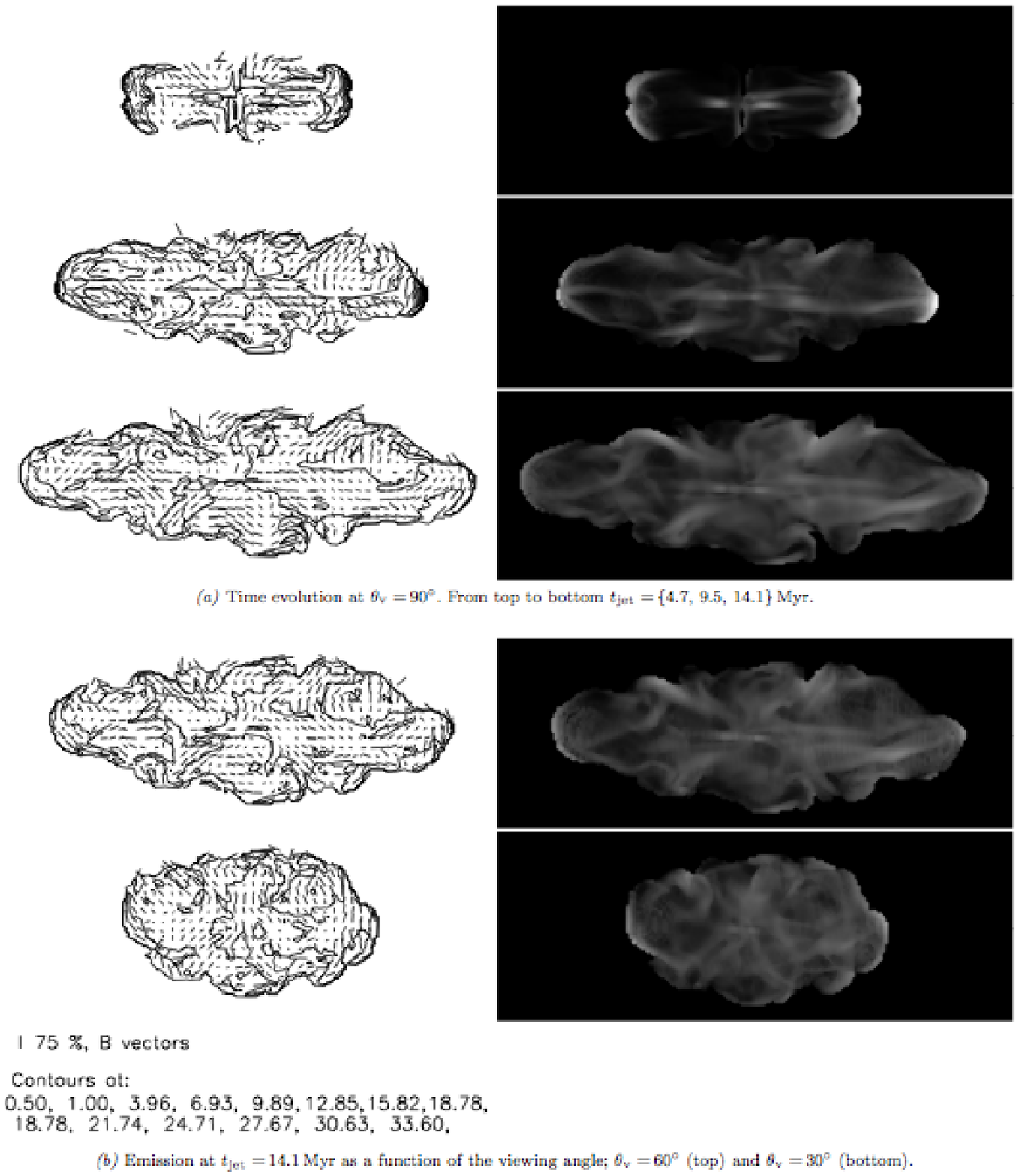}
  \caption{Synthetic observations of the source with \etaeq0.004 and Mach\,$=\,$40. \mapsCaptionA}
     \vspace*{0pt}
\label{pol-sim01-90deg-a}
\end{figure*}

\begin{figure*}
         \includegraphics[width=\textwidth]{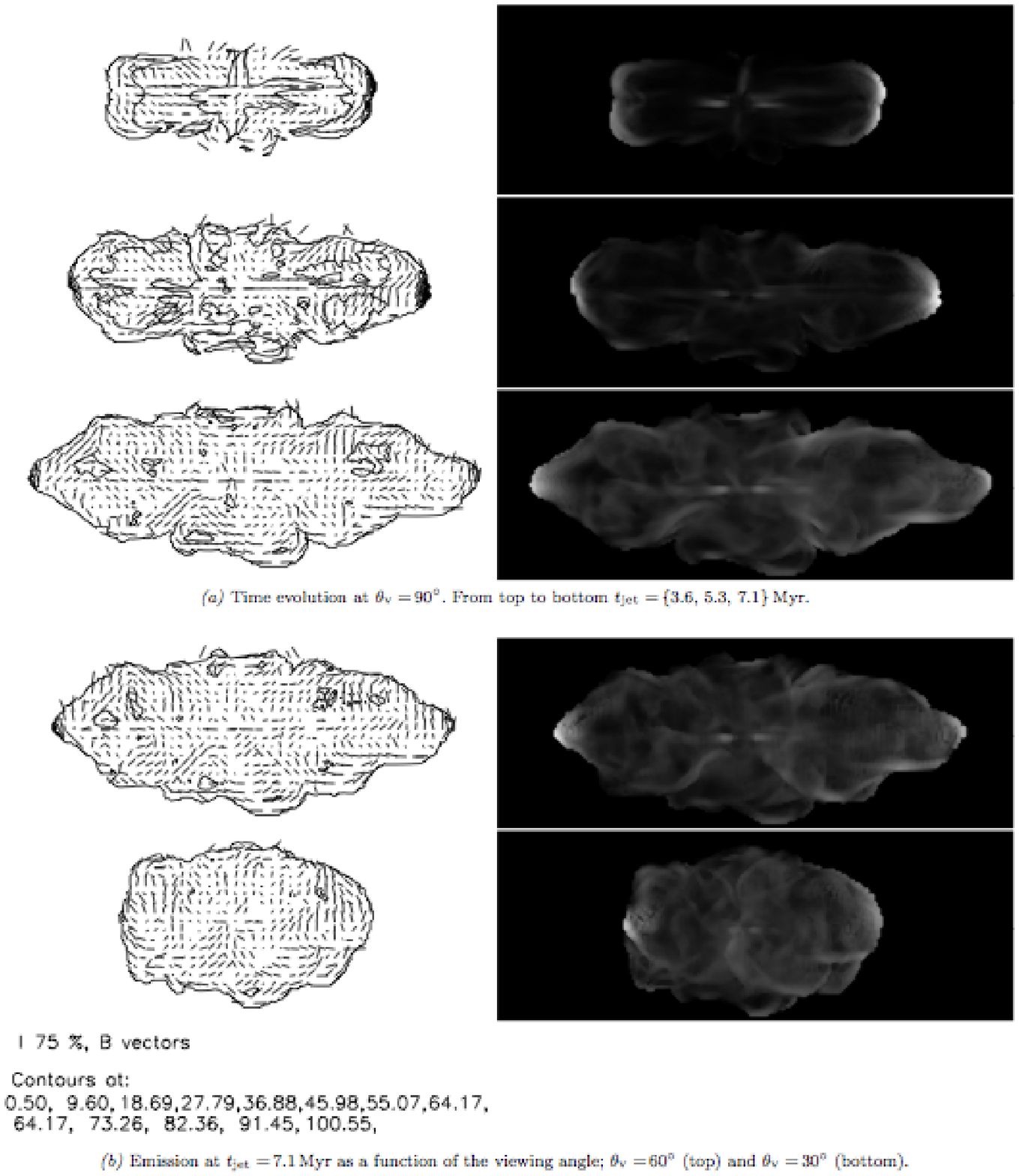}
  \caption{Synthetic observations of the source with \etaeq0.004 and Mach\,$=\,$80. \mapsCaptionA}
     \vspace*{0pt}
\label{pol-sim03-90deg-a}
\end{figure*}

\begin{figure*}
         \includegraphics[width=\textwidth]{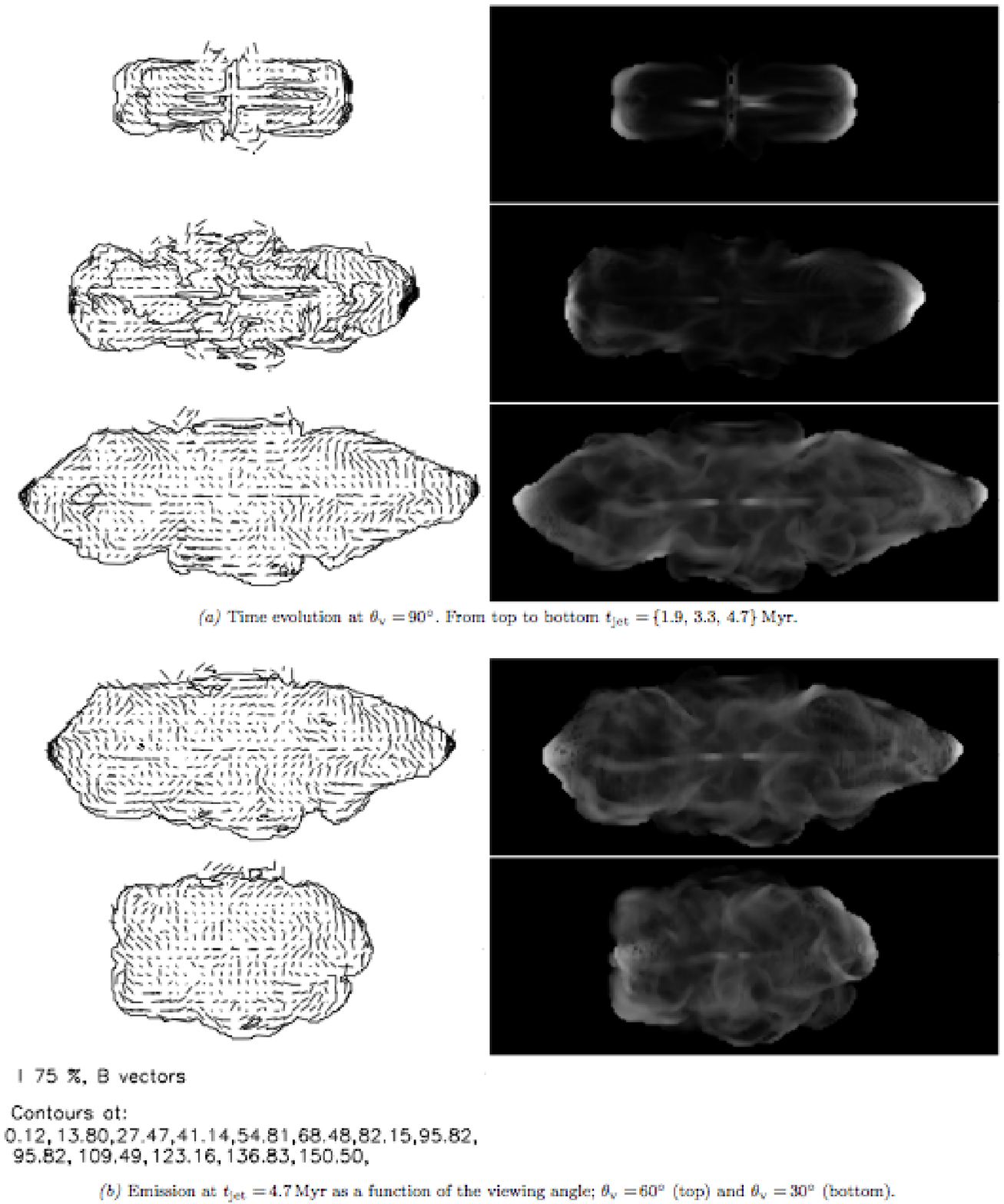}
  \caption{Synthetic observations of the source with \etaeq0.004 and Mach\,$=\,$130. \mapsCaptionA}
     \vspace*{0pt}
\label{pol-sim03-90deg-a}
\end{figure*}

\begin{figure*}
         \includegraphics[width=\textwidth]{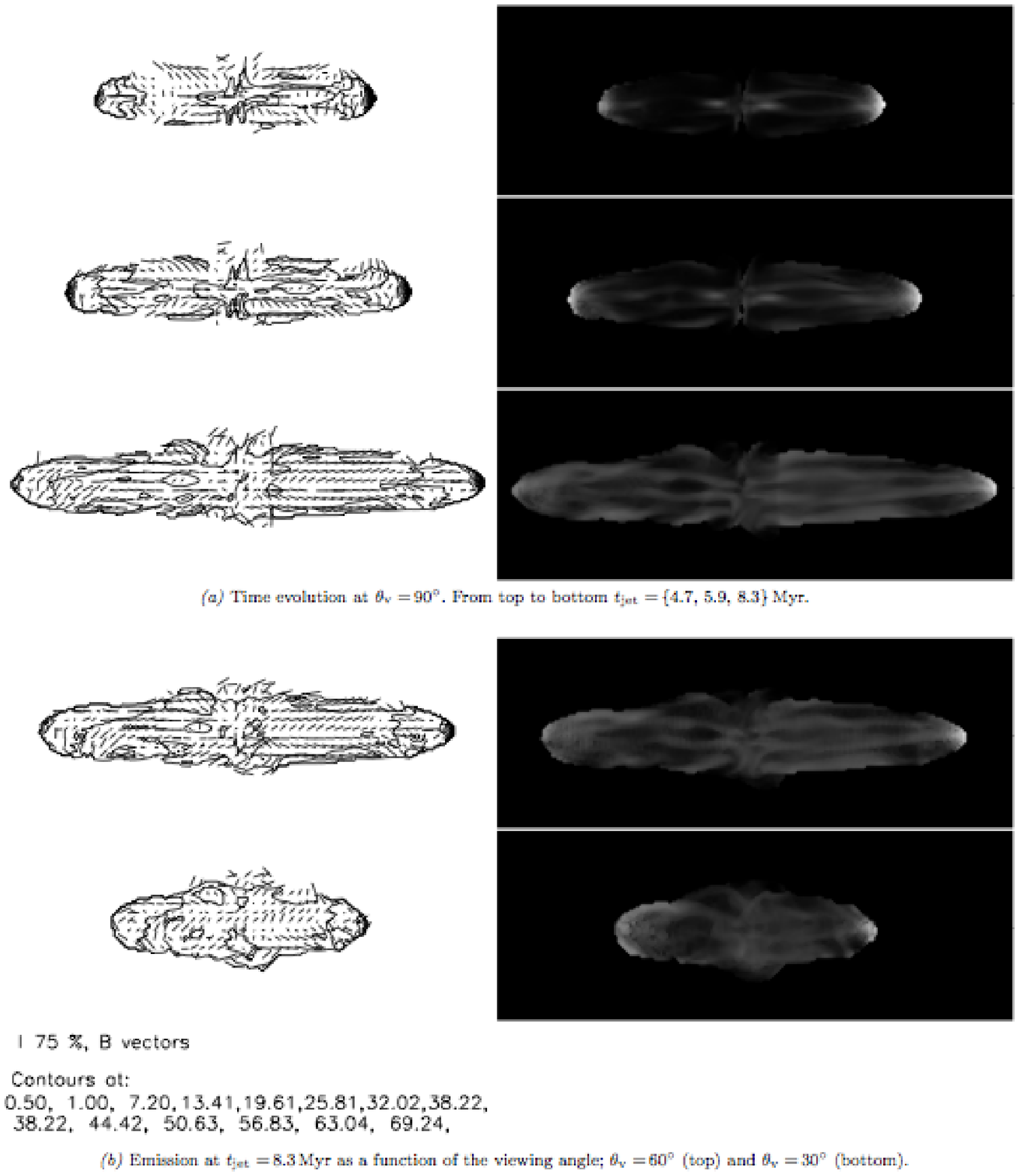}
  \caption{Synthetic observations of the source with \etaeq0.02 and Mach\,$=\,$40. \mapsCaptionA}
     \vspace*{0pt}
\label{pol-sim02-90deg-a}
\end{figure*}

\begin{figure*}
         \includegraphics[width=\textwidth]{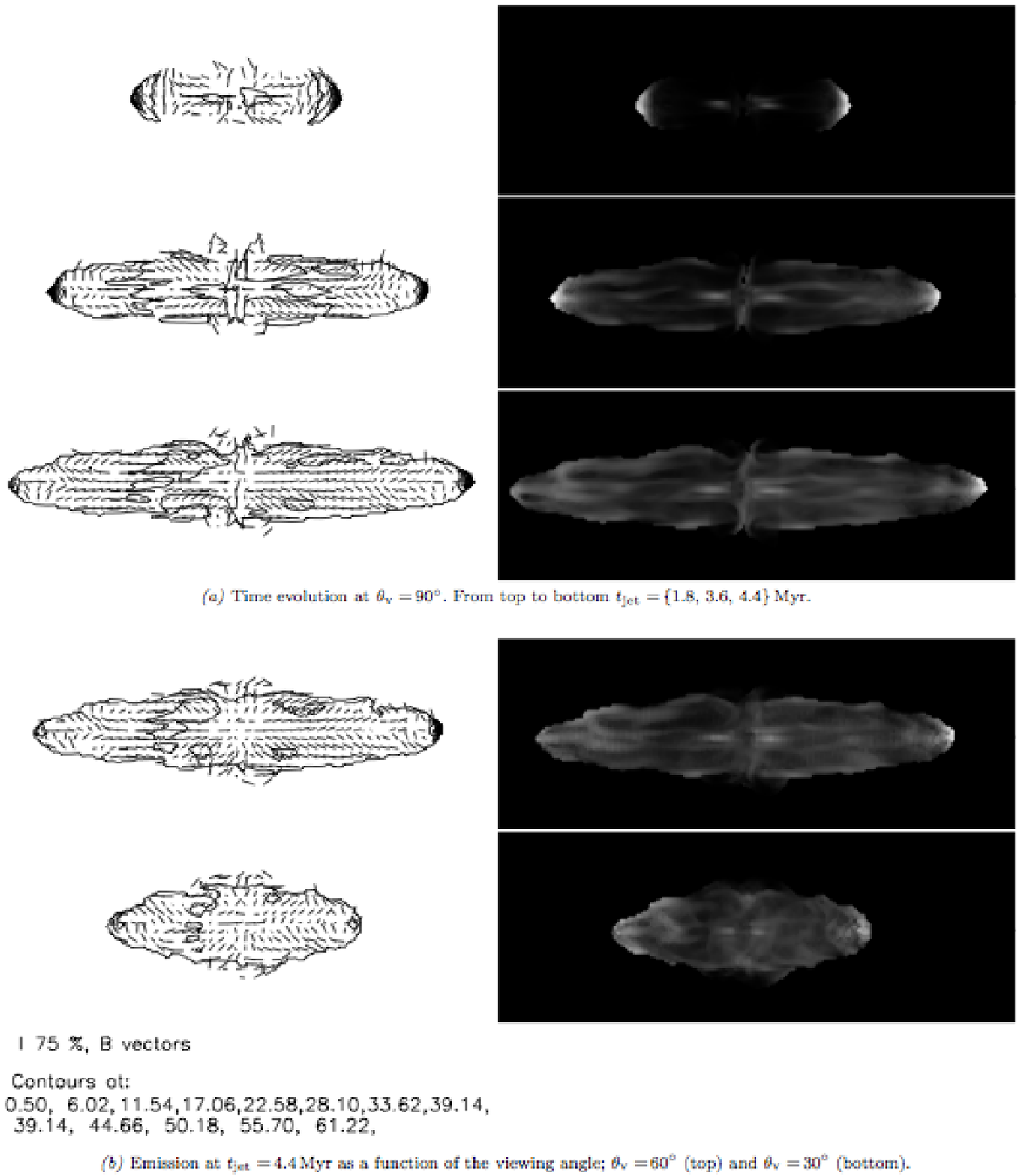}
\caption{Synthetic observations of the source with \etaeq0.02 and Mach\,$=\,$80. \mapsCaptionA}
     \vspace*{0pt}
\label{pol-sim04-90deg-a}
\end{figure*}

\subsection{Flow structure}\label{flow}

The hydrodynamic flow structure of our simulations is very similar to
what is generally found in the literature (compare
section~\ref{intro}, above, Figure~\ref{2dcuts}). The hypersonic jets flow
straight for a certain distance (2D-phase). Then three-dimensional
instabilities
develop, more clearly in the lighter jet runs, and the jet head
oscillates around the jet axis, consistently with
Scheuer's dentist drill (3D-phase). 
Cocoons are wider for lower
jet density and faster jets, as expected. The relatively heavier jets
($\eta=\,$2$\times$10$^{-2}$) 
propagate faster in 
the 
axial direction, and 
backflows in their cocoons are much less turbulent than in their
relatively lighter 
($\eta=\,$4$\times$10$^{-3}$) 
counterparts.

The evolution of cocoon magnetic fields is driven by the following
dynamics. The field is initially random inside the jet injection
volume. The injected momentum stretches field lines along the jet
direction. This puts energy into the axial field, which is therefore
amplified.  The other field components are simply advected out of
the injection volume, and their field strength drops with time,
within the injection volume and the beam.  This process results in
a poloidally dominated magnetic field, similar to the setup in
\citet{volker09}, yet with some important differences: First, the
field in the axial direction is patchy (compare Figure~\ref{2dcutsB}
for this and other details of the magnetic field), i.e. adjacent
parts of the beam have the field parallel and anti-parallel to the
flow direction. Second, for a given plane perpendicular to the flow
vector, the field may in principle also be patchy, i.e. there is
not necessarily one dominant toroidal field loop, but possibly two
or more field loops across a section of the jet. However, the fact
that the power spectrum used for the initial field setup strongly
favours larger scales, still produces a predominantly toroidal
configuration for the magnetic field perpendicular to the jet axis.

As Gaibler et al. (2009) 
do, we find that the axial field lines return to the
injection region very close to the jet. In our case this may even
happen inside the beam, since any beam cross section may in general 
contain axial field patches of opposing directions. In the presence of
a backflow, this requires field line reconnection, which should be
easily possible in the jet head on numerical grounds due the complex
flow pattern in this region. 
%
	This seems to suggest that the general structure of the
	poloidal magnetic field does at least not very much depend
	on the initial conditions. The reason is that the elongation
	of the beam stretches the axial field lines and therefore
	amplifies the axial field. To have the field lines going
	out in the beam and returning close to it or even within
	is the configuration which requires the least amount of
	energy, and is therefore chosen by the system.
   %
  
Gaibler et al. (2009) find
the toroidal part of the field, which cannot be lost to other field components
via turbulence due to the axisymmetry condition in their study, increases linearly
with distance from the jet axis. This may be easily understood from
the induction equation.
The physical reason is the work done by the expanding cocoon on the
toroidal field component is stored in that part of the magnetic field.
We do not observe such a linear increase in the toroidal field in our
3D simulations directly, but we expect this process also to be at
work. Since it is related to the cocoon expansion, we expect more
magnetic energy to be created 
by 
fatter cocoons, i.e. for lower
jet density (most important for the cocoon width)%
, 
and higher jet velocity, which is also found by
Gaibler et al. (2009). They also show that this process is able to enhance
the total magnetic energy in the jet by a factor of a few (see their
Figure~20). 
For our simulations, therefore, we expect
a noticeable increase in the magnetic
energy during the simulation time
; the fatter the cocoons the higher the energy rise.
Figure~\ref{cocoon-ene} shows 
this expectation is exactly what we
find: Here we plot the magnetic energy in 
both the cocoon and beam over the source
size for all runs. The curves are indeed strictly ordered according to
cocoon width. All the lighter jets have more magnetic energy than any
of the light ones. Among jets with a given density, the faster ones
have more magnetic energy. Therefore, the underlying
reason for the increase of the magnetic energy is the increase of the
toroidal component due to the cocoon expansion, just as in
Gaibler et al. (2009). There is another detail 
that 
confirms this
finding: As described above, we find the usual 2D and 3D-phases for our
jet simulations. The described amplification mechanism is very
different in 
each 
phase. During 
the 2D-phase, 
field loops released in the jet head expand axisymmetrically, and
substantial work is required to stretch the%
m. 
Energy from this work 
is later found in the magnetic field. 
In contrast, during 
the 3D-phase the dentist's drill moves the
jet head away from the axis
in different directions. 
Field loops therefore do not have to expand to reach
large distances from the axis. They may keep their size, and get 
pushed into different corners of the cocoon at different times. 
Hence, 
once cocoon inflation reaches and goes thought the 3D-phase,
almost no work is put into the field anymore.
We believe 
this mechanism causes the turnover in the magnetic energy seen in
Figure~\ref{cocoon-ene}. This turnover is visible for all the lighter
jets at the comparable source size.  The light jets%
, on the other hand, 
do not show much
of an amplification in the first place, and also remain quite
straight, i.e. essentially in the 2D-phase up to the end of the simulations. As
expected, we do not find the turnover there. We note that a similar
turnover is not found in the axisymmetric simulations by
Gaibler et al. (2009) either, which is of course expected if it is linked to the
3D-phase.

Why do we not see a linear increase with distance from the jet axis in the toroidal field like
Gaibler et al. (2009)? Because of the 3D nature of the cocoon
turbulence in our simulations. While axisymmetric turbulence can only
stretch and compress a given toroidal field, 3D-turbulence may also
turn toroidal field into poloidal one. The result is a turbulent
cocoon field, with no geometrical similarity to the 2D-result.

We see a strong axial field along the edge of the cocoons (see
Figure~\ref{2dcutsB}). This is due to velocity shear in this
region
%
  (Section~\ref{discu}) where the time average average backflow speeds with
  respect to the ambient medium are about 
  5$\times$10$^3$, 5$\times$10$^3$, 28$\times$10$^3$,
  22$\times$10$^3$ and 27$\times$10$^3$\,km\,s$^{-1}$, for the 
  sources as they appear in Table~1.
%

In the shocked ambient gas, on the other hand, magnetic fields are
first compressed in the bow shock, and then reduced again due to
adiabatic expansion of the gas, as it leaves the shock towards the
cocoon.  The effects of cocoon expansion on CMFs will be investigated
in a sequel paper 
(Huarte-Espinosa, Krause, \& Alexander 2011b, in prep.).

%
  The flow structure in our simulated radio sources is dominated
  by large scale motions, namely the toroidal and poloidal stretching
  mechanisms we have discussed in this paper. We cannot claim to
  represent the turbulence in our simulated cocoons well, because
  the resolution is too poor. Higher resolution should add additional
  small scale structure, unless prevented by a sufficiently strong
  magnetic field (compare e.g. \citealp{krause01}). Yet, also
  turbulence is expected to have most power on large scales.
  Therefore, while higher resolution studies will still be useful,
  we would expect the results discussed in this paper to hold.
%

\subsection{Synthetic radio maps}
\label{pol-obs}

In Figures~\ref{pol-sim01-90deg-a},~\ref{pol-sim03-90deg-a},~\ref{pol-sim02-90deg-a}
and~\ref{pol-sim04-90deg-a},
we present synthetic radio 
and polarisation angle maps of four of
our simulated FR~II radio sources 
(all but the lighter-relativistic one)
for three different snapshot times.
These maps essentially reflect the field
structure discussed above.
The emission is dominated by filaments, hotspots and
sometimes jets are seen. This is 
similar to what has been
found in earlier studies, as detailed in the Introduction.
The jet head region is more prominent at earlier times and for higher
jet density. Our lighter jets feature more diffuse jet heads
reminiscent of the {\em shock web complex}, described by \citet{
 treguillis01}.
Our polarisation angle maps are all dominated by larger patches. This
is due to the fact that the cocoon dynamics is dominated by large
vortices, about the cocoon radius in diameter. The backflow in our
$\eta=0.02$ 
jets remains quite smooth and, consequently, the
polarisation vectors are even more parallel to the jet axis than they are
in the lighter~$\eta$ sources. 
Generally, we find an almost one-to-one
correspondence between the flow field and the polarisation vectors, as
expected.

In addition, in panel~(b) of
Figures~\ref{pol-sim01-90deg-a},~\ref{pol-sim03-90deg-a},
\ref{pol-sim02-90deg-a} and~\ref{pol-sim04-90deg-a} we show synthetic
radio and polarisation angle maps at different viewing angles.  We
consistently see that at small angles the axial field component
gets smaller due to the projection effect, while other field
components become relatively more prominent.

The patchy distributions in our polarisation maps are in good
agreement with typical observations of FR~II sources and radio-loud
quasars \citep[see e.g.][]{bridle84,saikia88,gilbert04,mullin06}. Along
the projected direction of 
jets we see that \pangle$<\,$20\de, which
are smaller angles than elsewhere inside the cocoon.  For 
\etaeq0.02, 
\pangle\ increases progressively 
along the vertical direction, from the jet axis to the edge of
sources
(Figures~\ref{pol-sim02-90deg-a} and ~\ref{pol-sim04-90deg-a}). 
This is because in these simulations cocoons are narrow, and therefore the
beam contributes significantly to the emission, which is not
the case in most of the observed sources.
In contrast, for 
\hbox{$\eta=\,$0.004}, 
\pangle\ shows
weak trends along the vertical direction.  The outermost vectors
in all the maps are commonly tangent to the dimmest emission contours.
This is similar to observations, but is of course influenced by 
numerical problems at the contact surface, as outlined above.

Polarisation degrees within \hbox{37--51\,\%} are found inside
cocoons, but higher, up to~$\sim\,$63\%, both at the edge of sources and 
at the position of jets.  We often see uniform patches with very
similar values of \pangle\ and $p$ at the
position of bright emission shocks. The vectors frequently follow
strong intensity gradients perpendicularly and have $p \ga \,$50\%.
Regions of non-uniform \pangle, on the other hand, are frequently
located between emission shocks (see e.g.  Figure~\ref{pol-sim01-90deg-a}).
These correlations are in good agreement with observations (e.g. see
\citealp{hogbom79,laing81a,bridle84,saikia88,hardcastle97,leahy97,gilbert04})
and with models of plasma compression and shear
\citep[e.g.][]{laing81b,miller85}.

Our synthetic emission maps often show hotspots at the location of the
jets' working surfaces as well as filaments in the radio lobes.
Radio hotspots and filaments are often seen in well resolved FR~II
sources \citep[e.g. Cygnus~A,][]{perley84}.

We see the backflow of the (anti-parallel) jets collide and form
sheets near the cocoon equatorial plane (the one normal to the
jets and containing the central engine).  There, our polarisation
maps show B-vectors with $|\chi_\mathrm{B}|>\pi/$4 above and below
the centre of Figures~\ref{pol-sim01-90deg-a}--\ref{pol-sim04-90deg-a}
(left column).  We found instances of such polarisation angle
distributions in the observations of 3C~34, 3C~336 and 3C~341
(\citealp{johnson95}, \citealp{mullin06} and \citealp{gilbert04},
respectively).

   \begin{figure*}
     \begin{center}
		 \includegraphics[width=\textwidth]{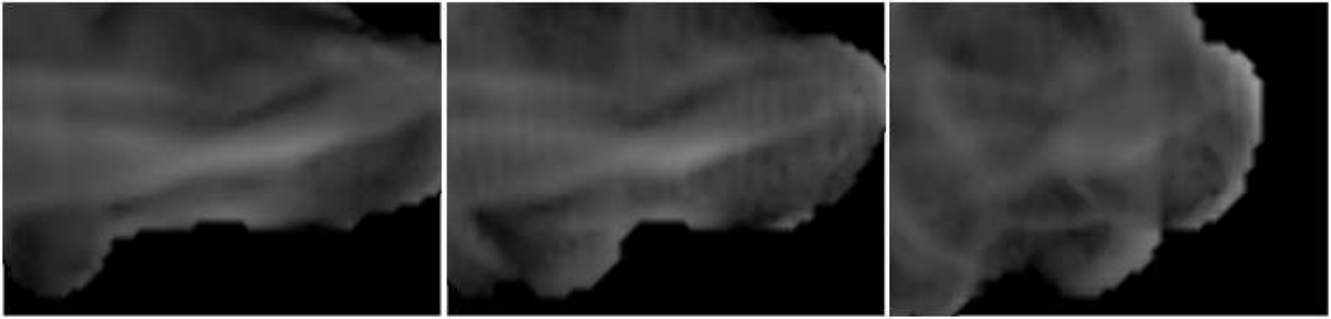} 
     \end{center}
     \caption{Synthetic emission filaments in the right lobe of the source 
with $\eta=\,$0.004
and Mach$\,=\,$40, at $t_{jet}=\,$14.1\,Myr. 
The structure 
at the centre of the figures 
gets shorter and dimmer as the
viewing angle decreases from 90~to 30~degrees, from left to right, 
suggesting a tube-like geometry for the feature in question.
}
   \label{filaments}
   \end{figure*}

At the end of the simulations we see 
laminar flows in the cocoons and also that both 
Rayleigh-Taylor and Kelvin-Helmholtz instabilities are
growing at the contact surface (Figure~\ref{2dcuts}). 
  Such flows form tube-like structures or filaments, as we see 
  in our synthetic emission maps 
  (Figures~\ref{pol-sim01-90deg-a}-\ref{pol-sim04-90deg-b}). 
Figure~\ref{filaments} shows
emission maps of the right lobe of the model source with $\eta=\,$0.004, 
Mach$\,=\,$40 and $t_{jet}=\,$14.1\,Myr, 
for viewing angles of 30, 60
and 90\,degrees. It is clear the structure 
at the centre of the figures 
gets shorter and dimmer as the viewing angle decreases.
This suggests a tube-like geometry for this feature.
  All our synthetic emission maps show filaments, however,
  we note that they form earlier in the low-$\eta$ sources than
  in the high-$\eta$ ones.

\subsection{Polarimetry and statistics.}
\label{pol-vs-eta}

In order to analyse our synthetic observations we have produced histograms
of the polarisation angle and the degree of linear polarisation;
see 
Figures~\ref{histoArrayAngle} and~\ref{histoArrayDegree},
respectively. 
In what follows we will see that 
these distributions show a clear correlation
with the viewing angle, the jet-to-ambient density contrast and time too,
but only a weak dependence on the jet velocity.

\subsubsection{The role of the viewing angle.}
\label{pol-vs-view}

  The polarisation angle histograms are similar for all runs at
  $\theta_\mathrm{v}=\,$90\de: They are all peaked towards zero
  degree, corresponding to the jet direction. The more isotropic
  distribution at lower viewing angle is consistent with cocoon
  turbulence. 
     We see only the distribution of the heavier jets remains peaked
     at a viewing angle of $60^\circ$, because of the weaker cocoon
     turbulence in these sources, relative to the ones with lighter
     jets.
  This confirms 
  the magnetic field structure is determined by the relative
  importance of turbulence 
  as well as 
  the amplification of the axial field
  due to the backflow in 
  cocoons. 

  For 
  $\theta_\mathrm{v}=\,$30\de, 60\de\ and 90\de, 
  the mean value of 
  $|\chi_\mathrm{B}(\eta=\,$0.004)$|$ 
  is generally
  larger 
  than that of 
  $|\chi_\mathrm{B}(\eta=\,$0.02)$|$ 
  (see Section~\ref{pol-vs-eta}, below). 
  The dispersion of the polarisation angle seems to follow this trend as well.
  The differences are
  pronounced for viewing angles of 60~and 90~degrees and related
  to the size of the data sample. i.e. the cocoons' volume, which
  is inversely proportional to $\eta$ in a non-linear way
  (Section~\ref{flow}). 
Polarisation angle histograms at 
$\theta_\mathrm{v}=\,$30\de\ 
show both the flattest gradients and the least number of vectors
amongst all histograms, 
and their distribution does not show 
a Gaussian functional form. 

\begin{figure*}
     \begin{center}
~~~~~~~~~~~~~~~~~~~~~~~~~~~~~~~
{\large $\theta_\mathrm{v}=\,$30$^{\circ}$}
~~~~~~~~~~~~~~~~~~~~~~~~~~~~
{\large $\theta_\mathrm{v}=\,$60$^{\circ}$}
~~~~~~~~~~~~~~~~~~~~~~~~~~~~~
{\large $\theta_\mathrm{v}=\,$90$^{\circ}$}
~~~~~~~~~~~~~~~~~~~~~~~~~~ \\
\vskip.1cm
\includegraphics[width=0.302\textwidth,bb=40 75 470 395,clip=]
  {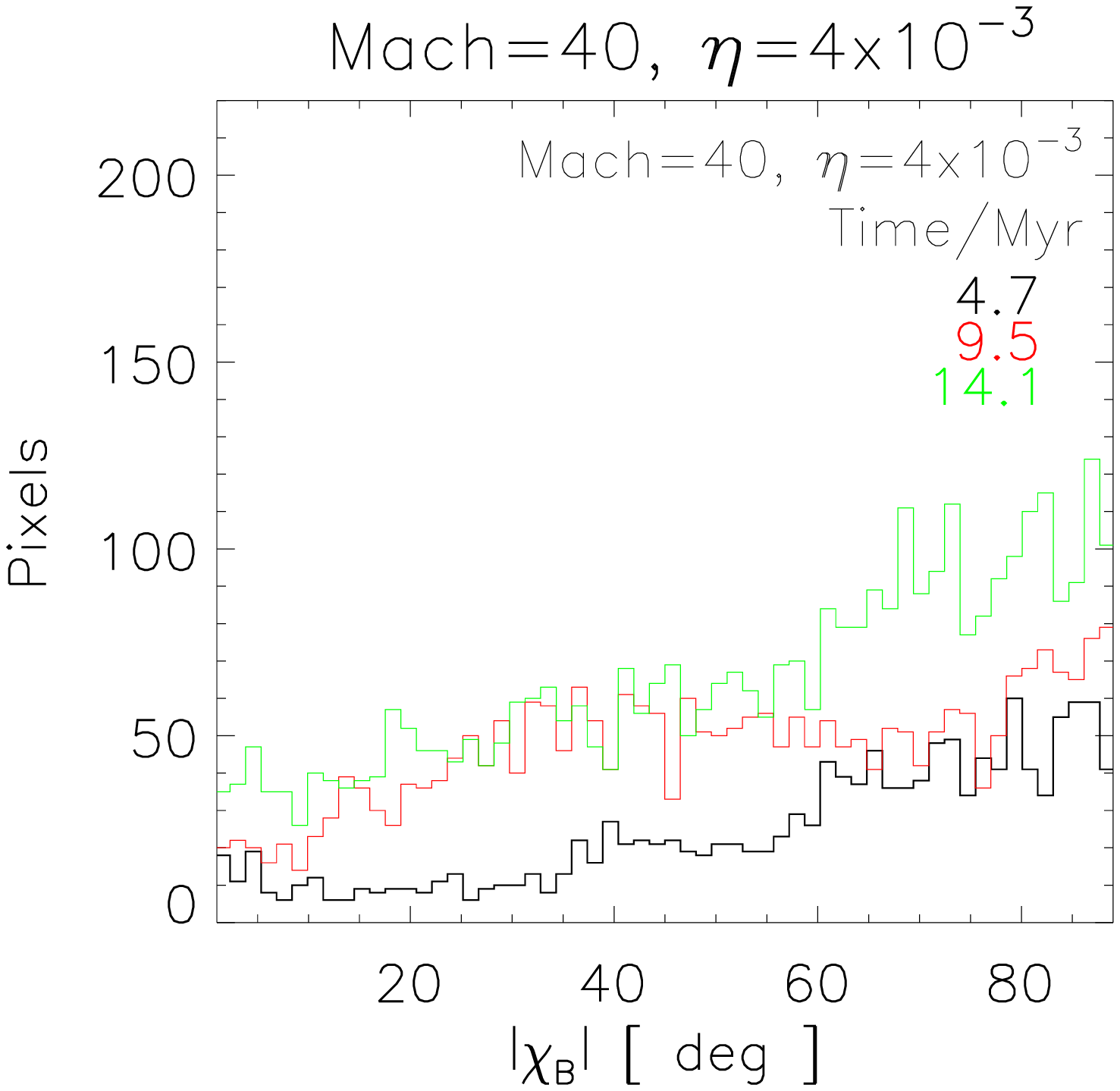}
\hskip-.21cm
\includegraphics[width=0.245\textwidth,bb=121 75 470 395,clip=]
  {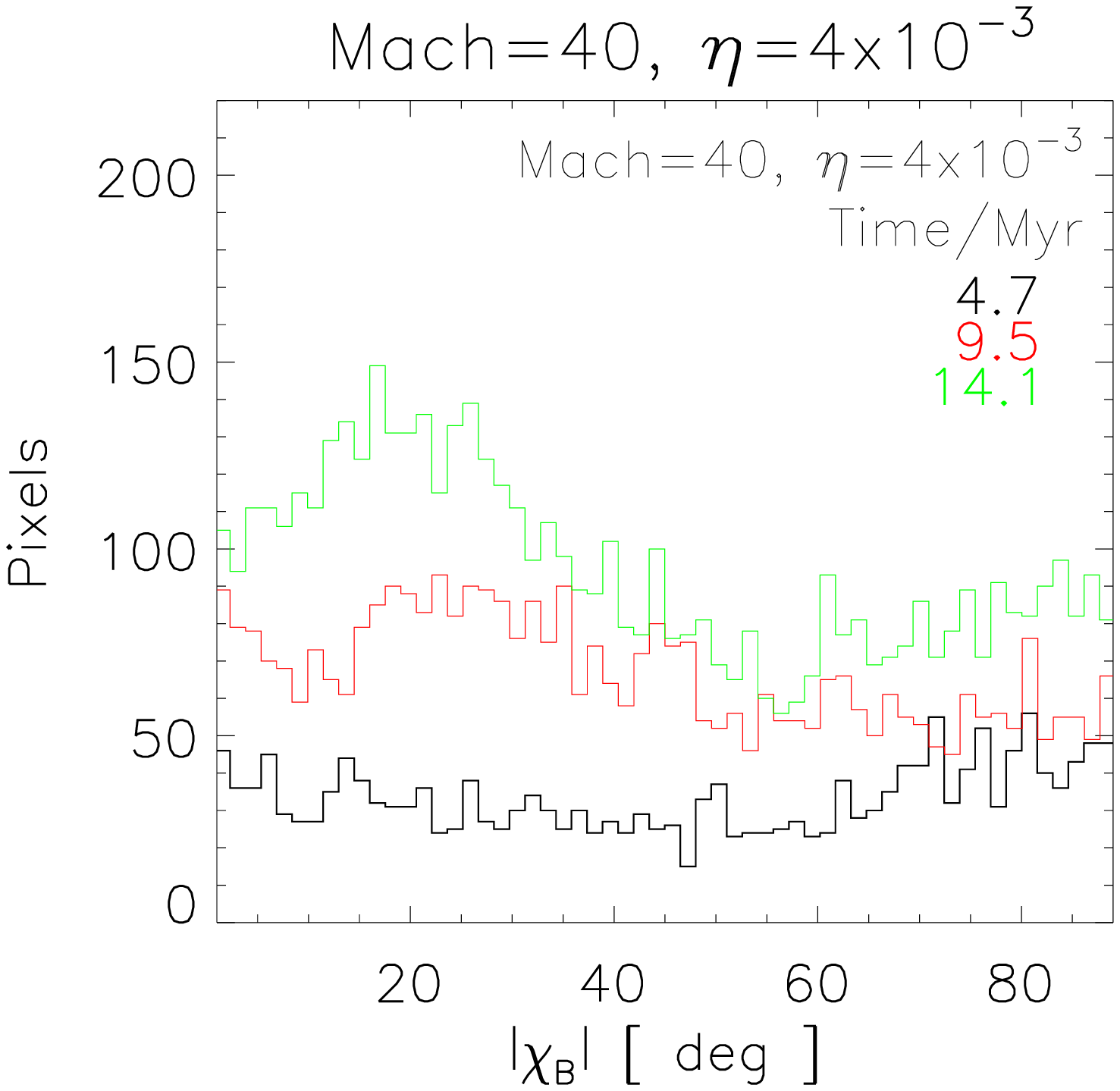}
\hskip-.21cm
\includegraphics[width=0.245\textwidth,bb=121 75 470 395,clip=]
  {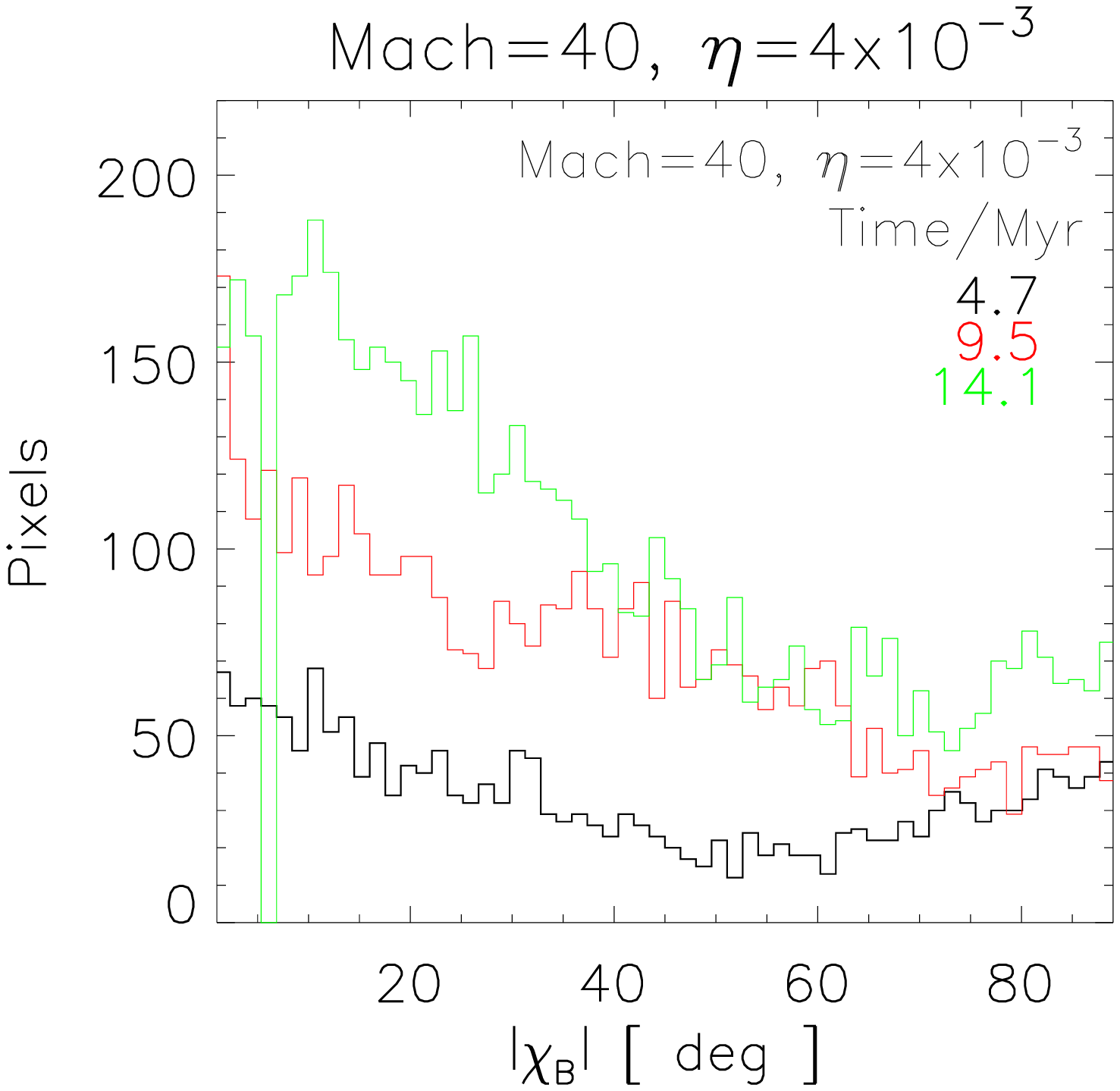} 
%
\vskip-.13cm
\includegraphics[width=0.302\textwidth,bb=40 75 470 395,clip=]
  {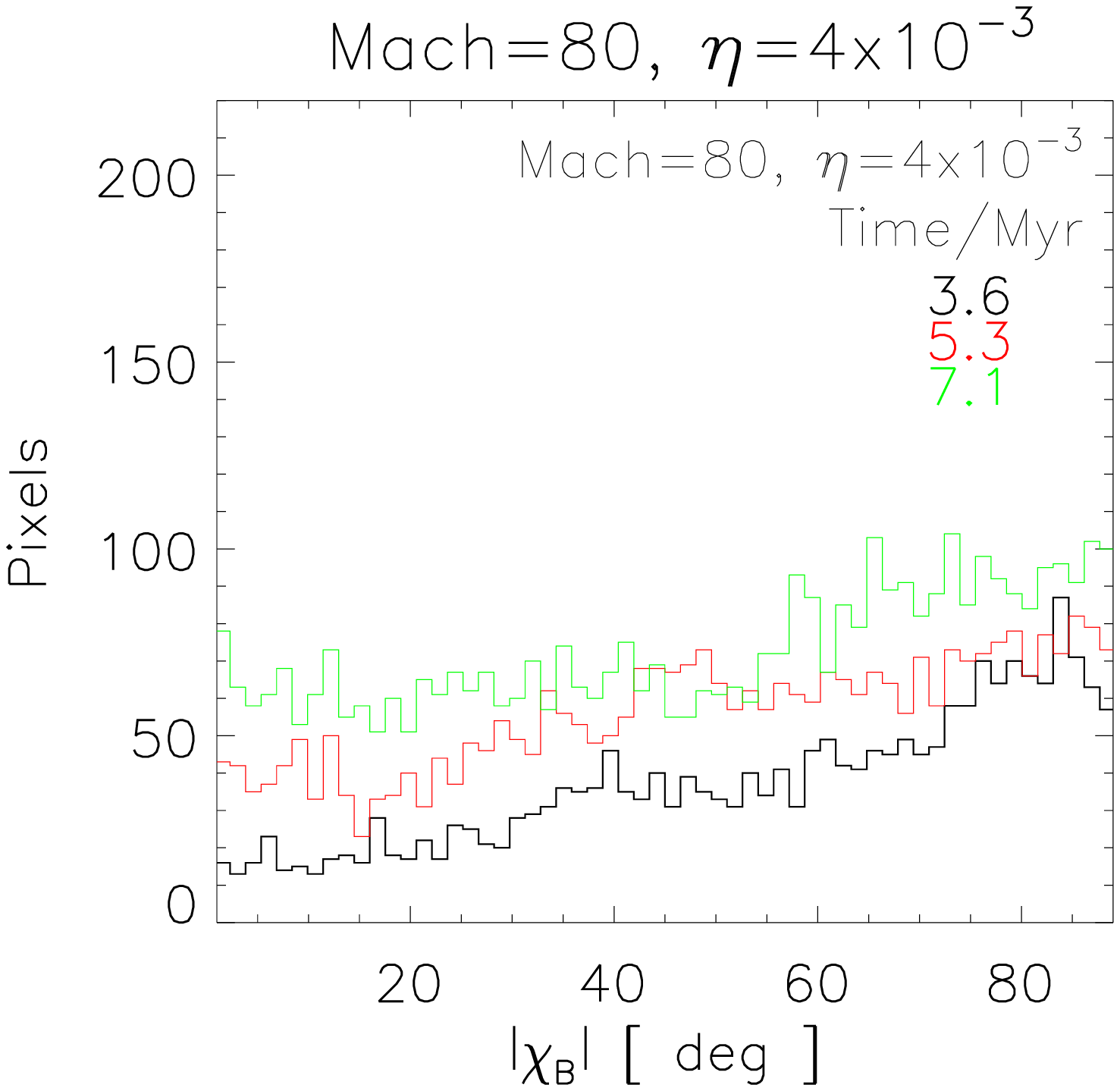}
\hskip-.21cm
\includegraphics[width=0.245\textwidth,bb=121 75 470 395,clip=]
  {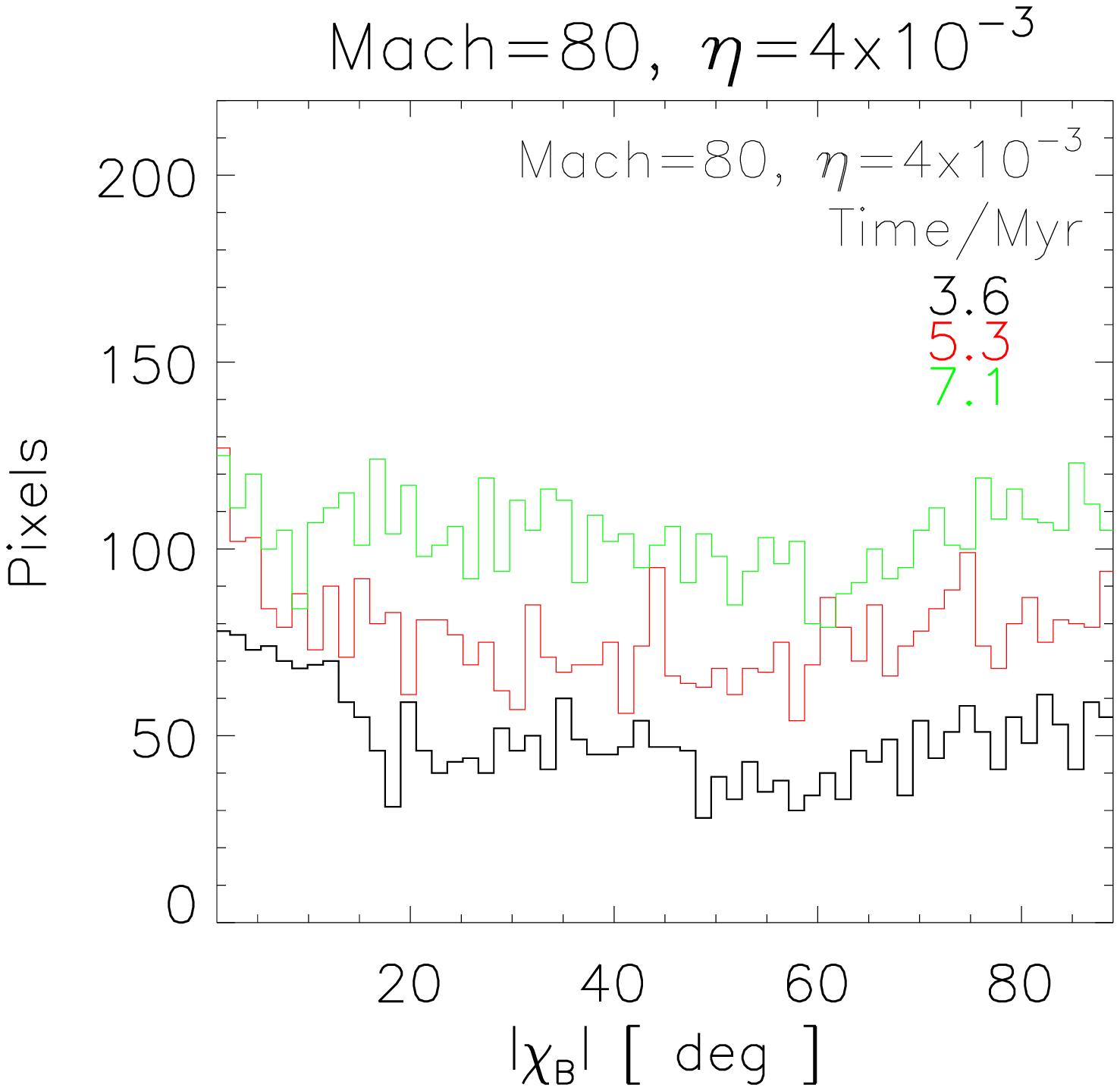}
\hskip-.21cm
\includegraphics[width=0.245\textwidth,bb=121 75 470 395,clip=]
  {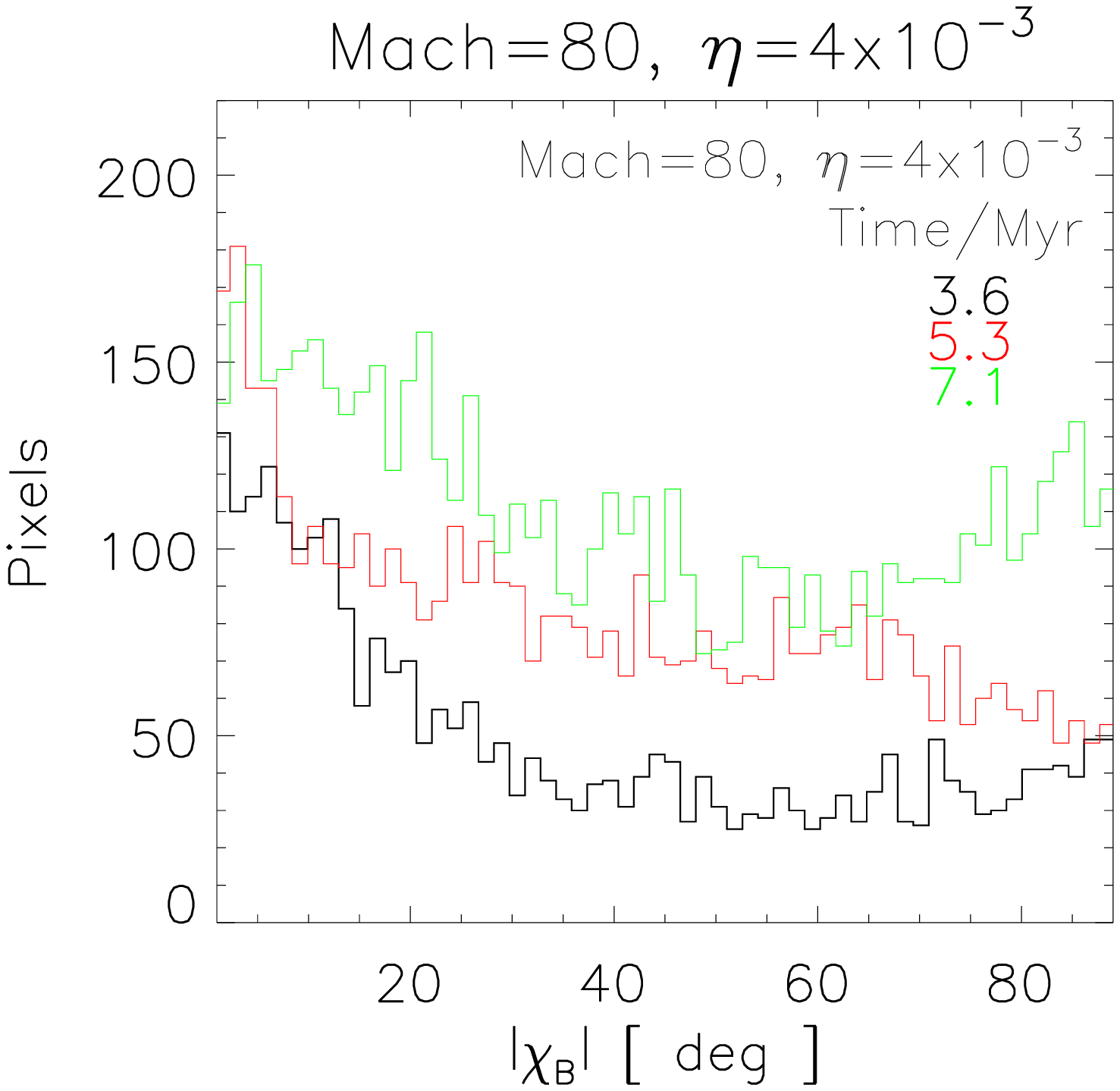}
%
   \vskip-.13cm
   \includegraphics[width=0.302\textwidth,bb=40 75 470 395,clip=]
     {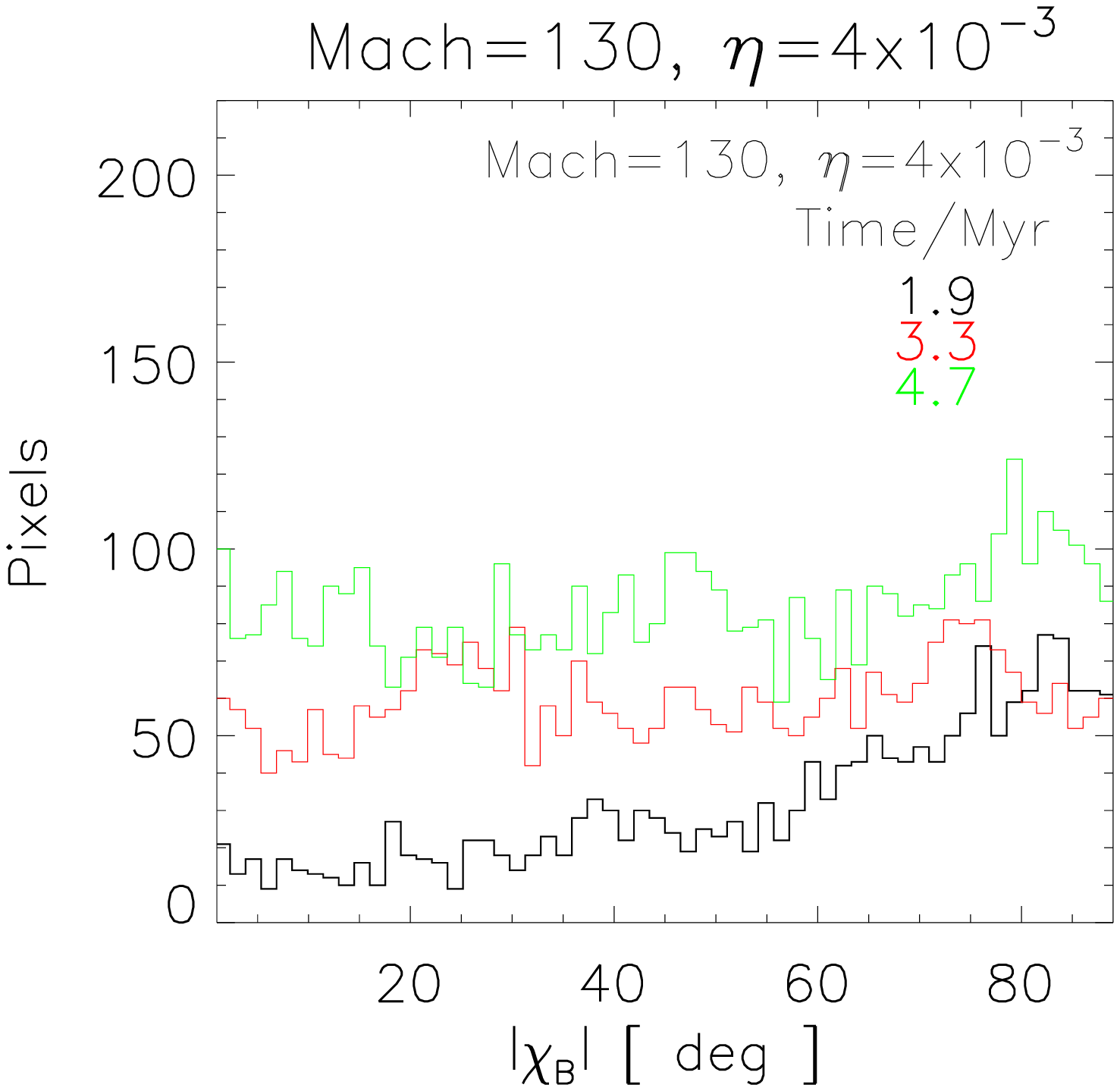}
   \hskip-.21cm
   \includegraphics[width=0.245\textwidth,bb=121 75 470 395,clip=]
     {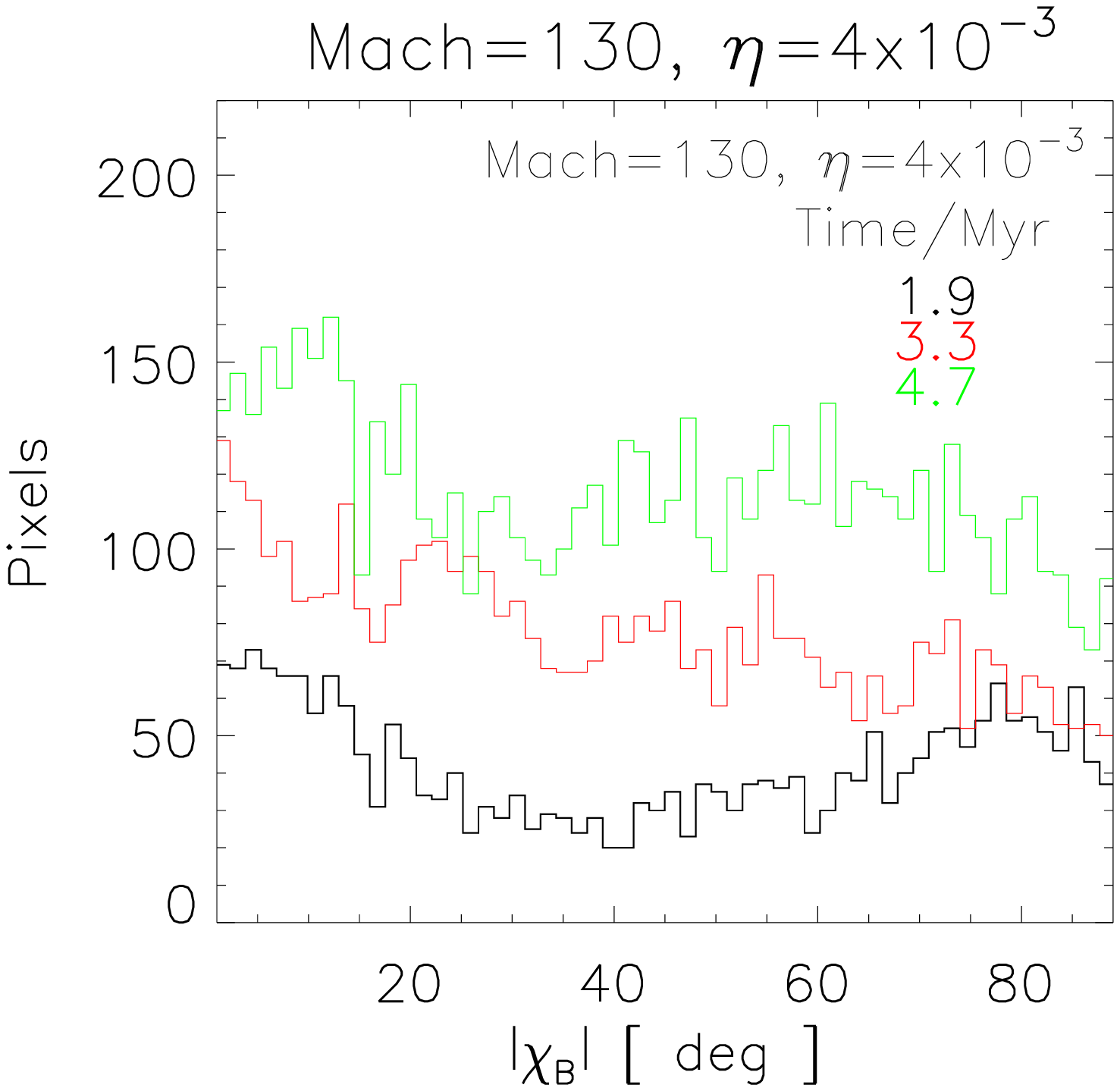}
   \hskip-.21cm
   \includegraphics[width=0.245\textwidth,bb=121 75 470 395,clip=]
     {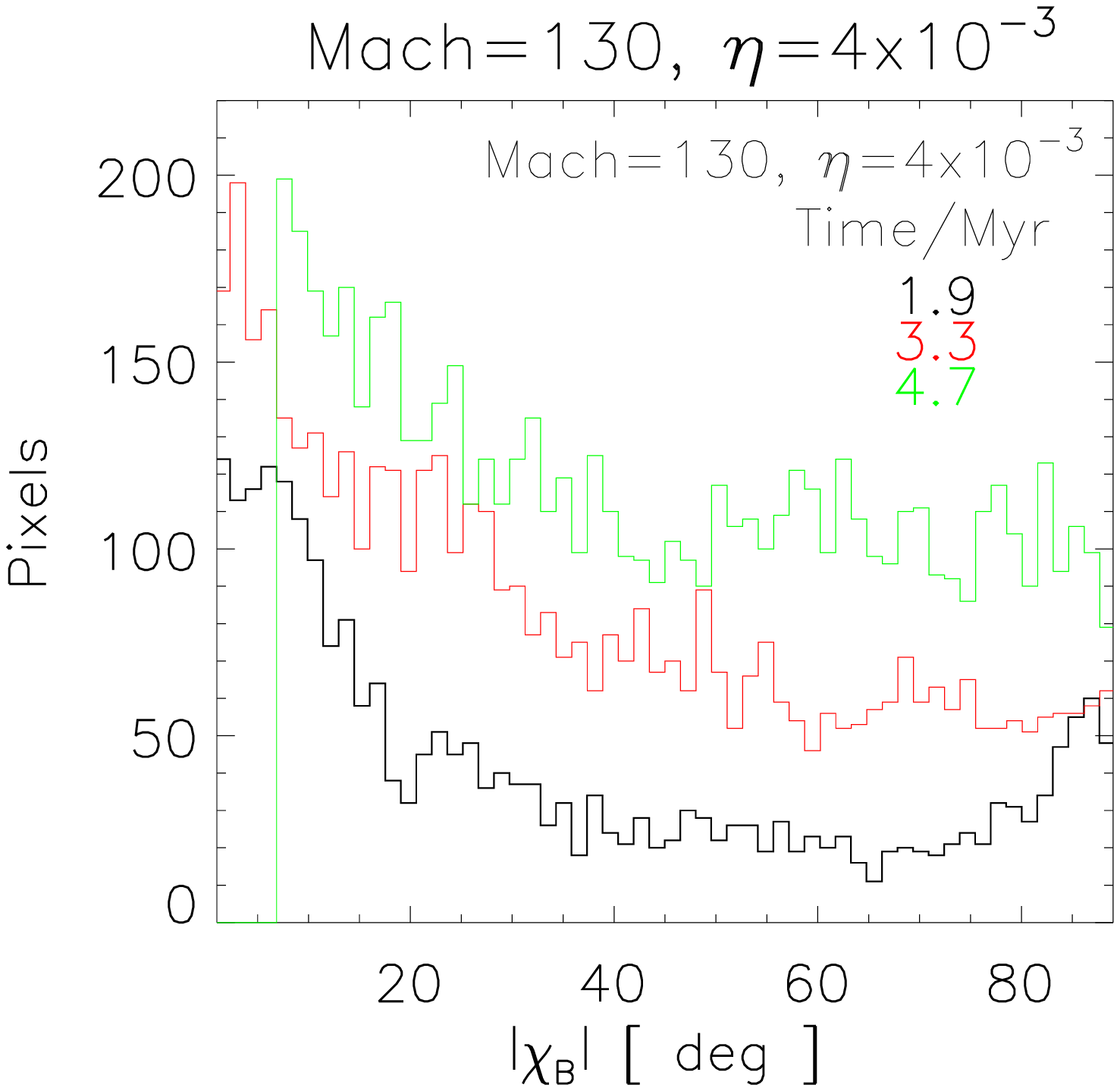}
   %
\vskip-.13cm
\includegraphics[width=0.302\textwidth,bb=40 75 470 395,clip=]
  {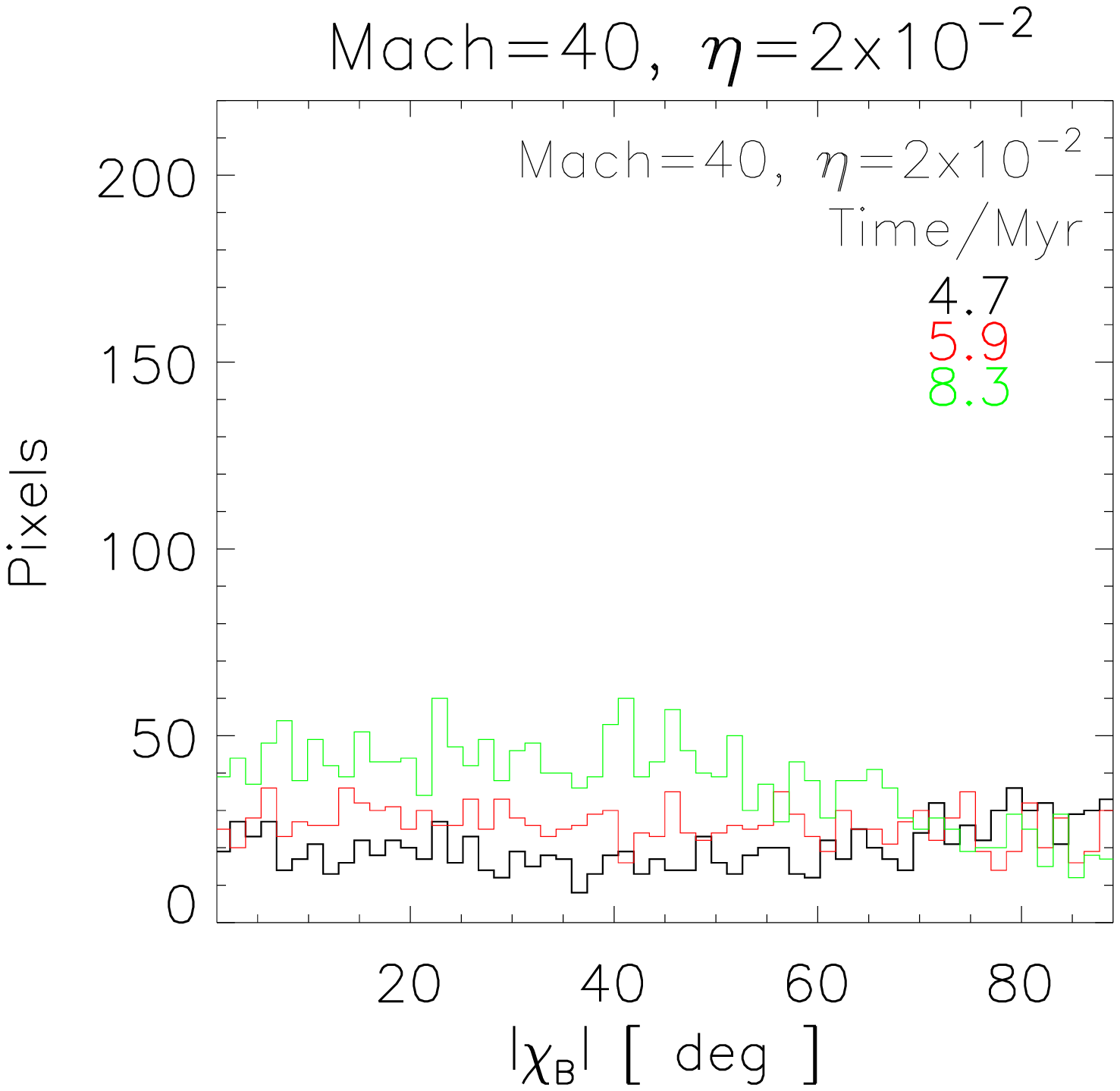}
\hskip-.21cm
\includegraphics[width=0.245\textwidth,bb=121 75 470 395,clip=]
  {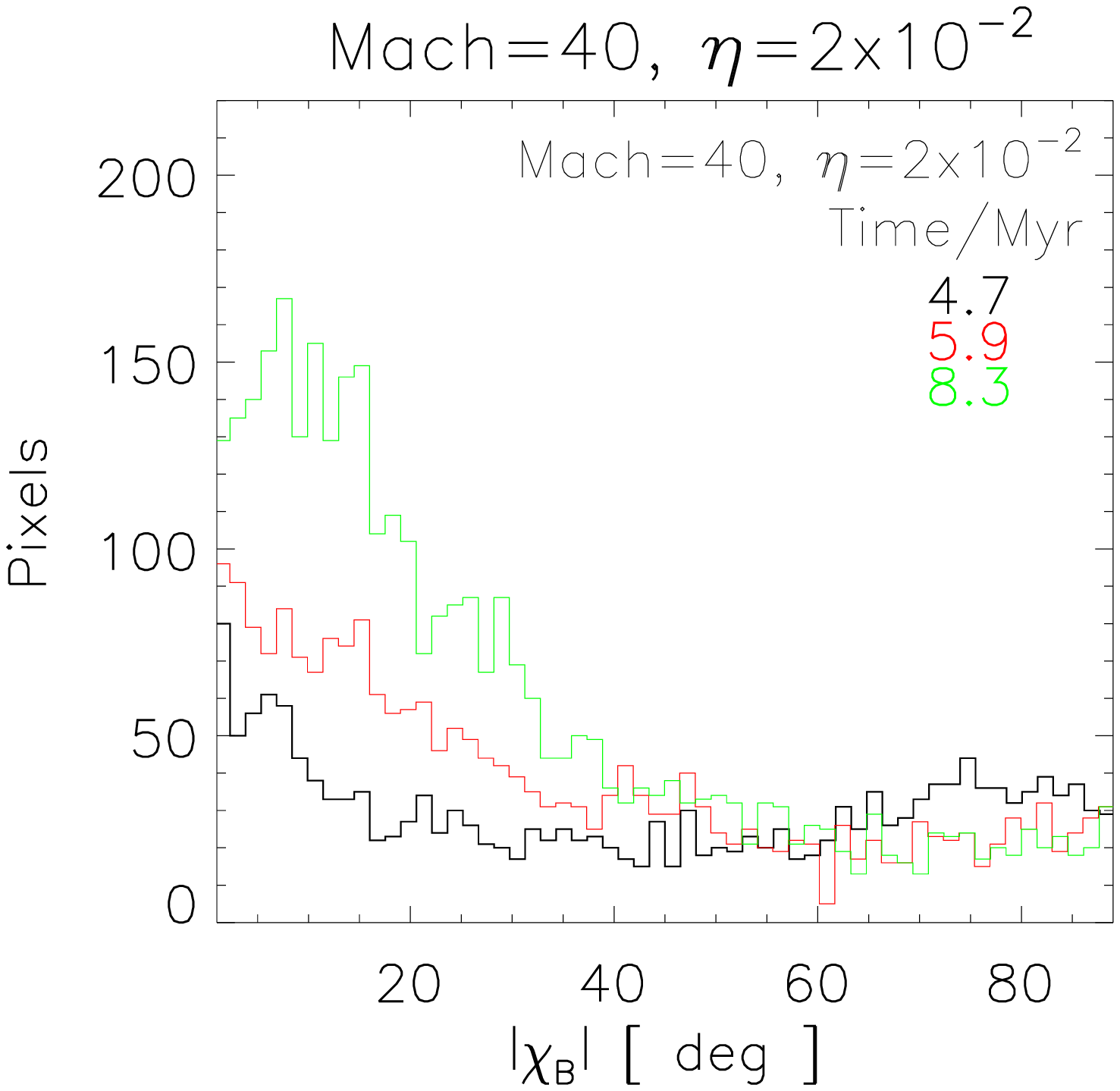}
\hskip-.21cm
\includegraphics[width=0.245\textwidth,bb=121 75 470 395,clip=]
  {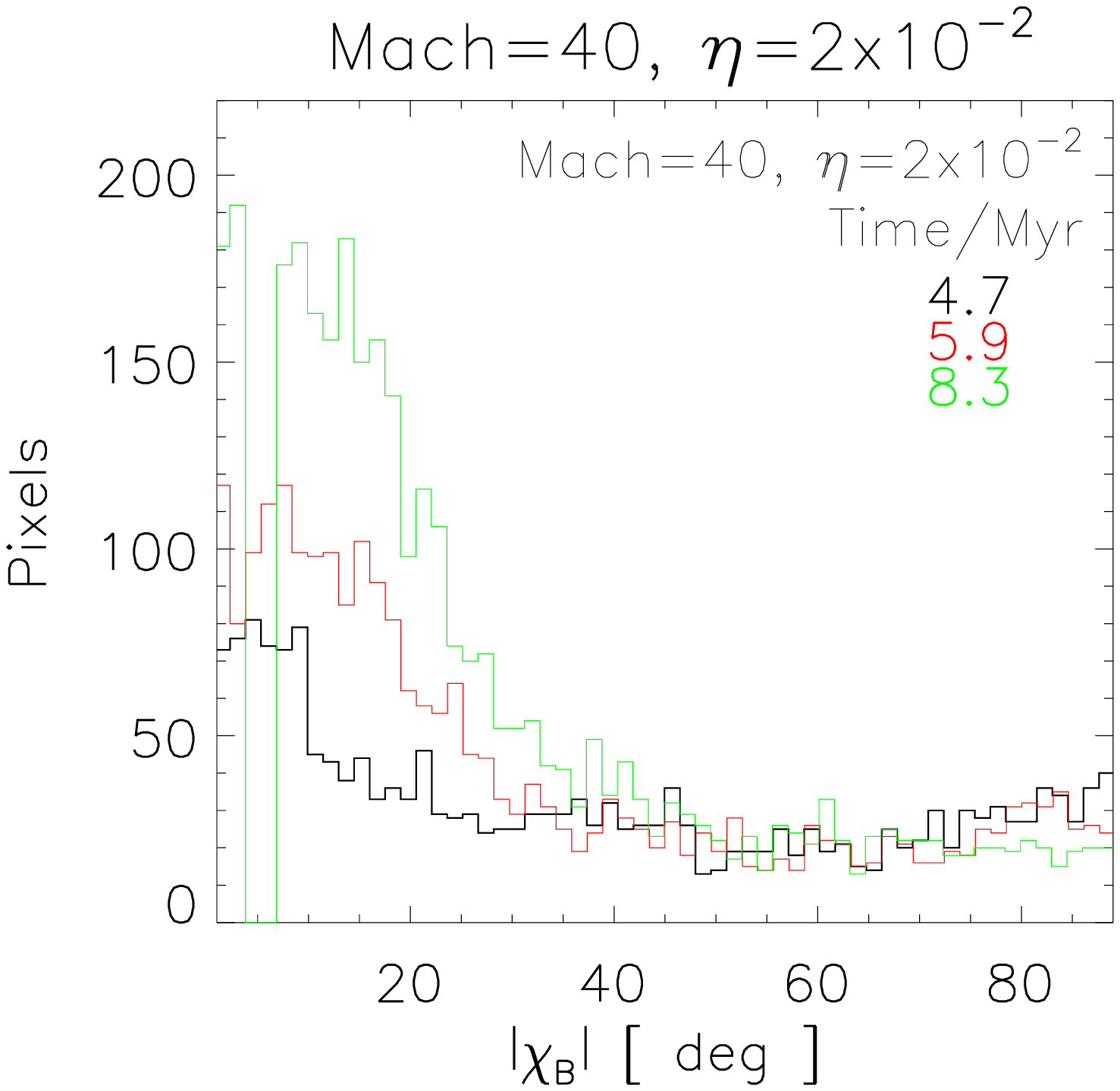}
%
\vskip-.13cm
\includegraphics[width=0.302\textwidth,bb=40 10 470 395,clip=]
  {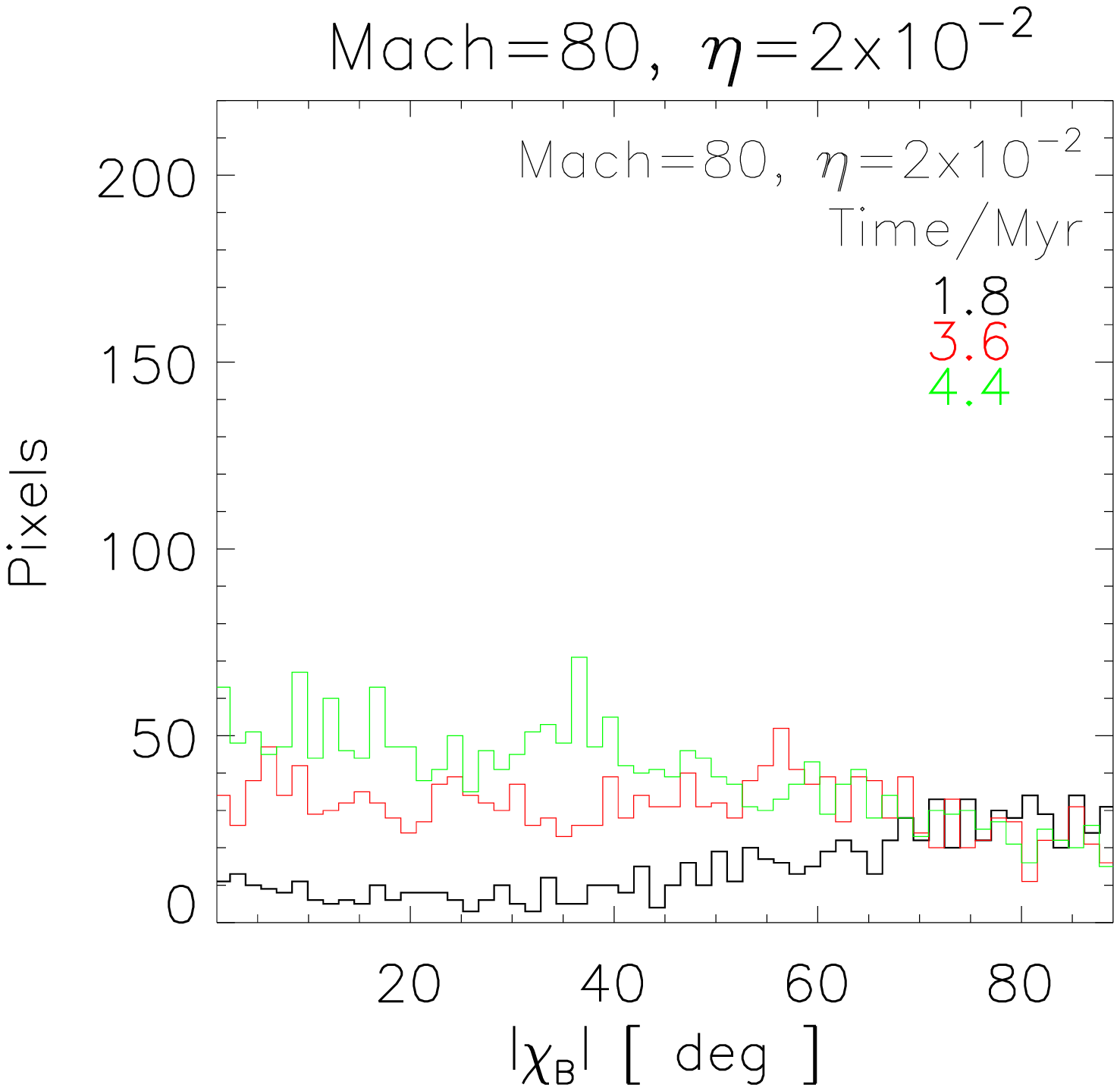}
\hskip-.21cm
\includegraphics[width=0.245\textwidth,bb=121 10 470 395,clip=]
  {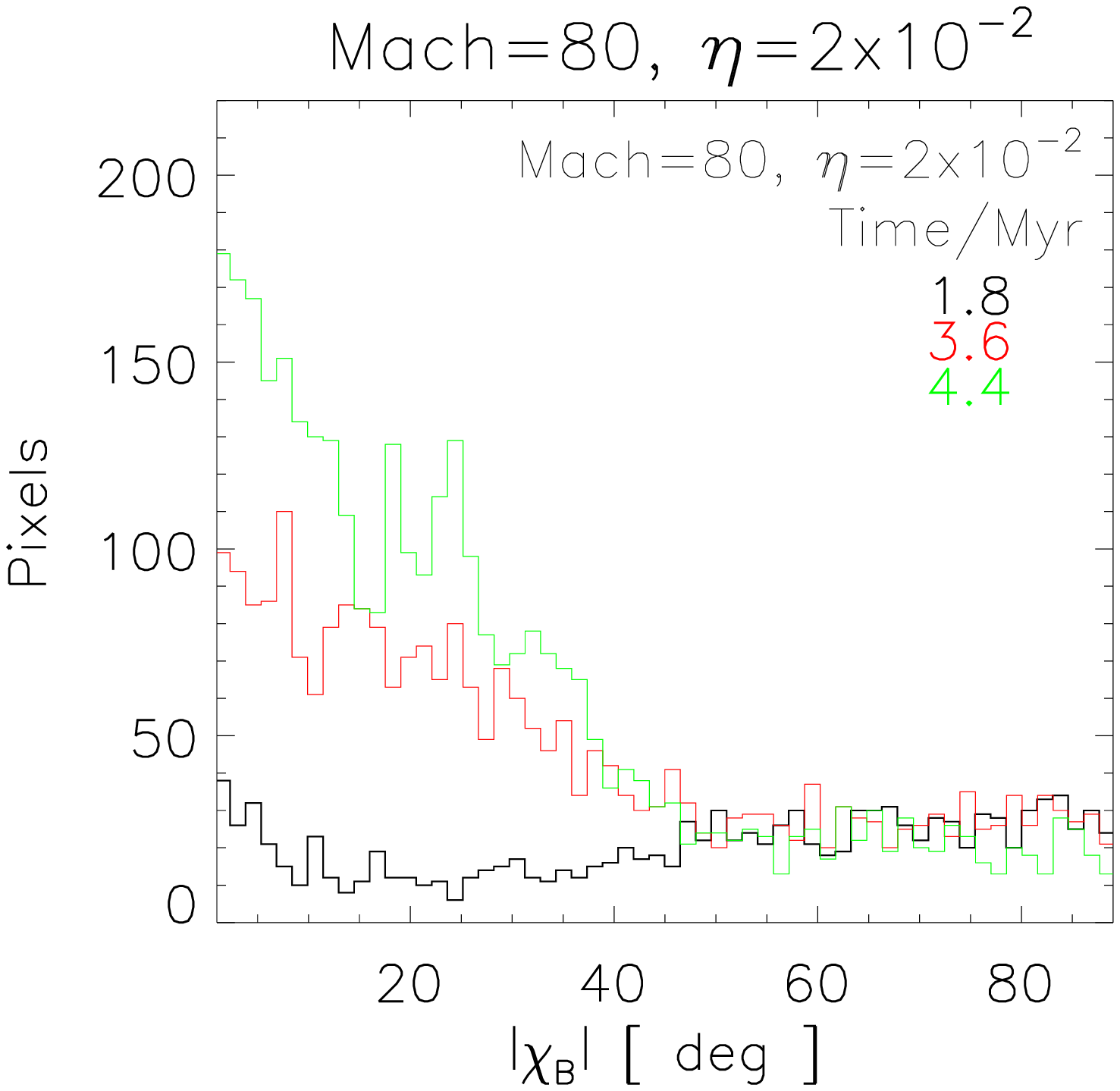}
\hskip-.21cm
\includegraphics[width=0.245\textwidth,bb=121 10 470 395,clip=]
  {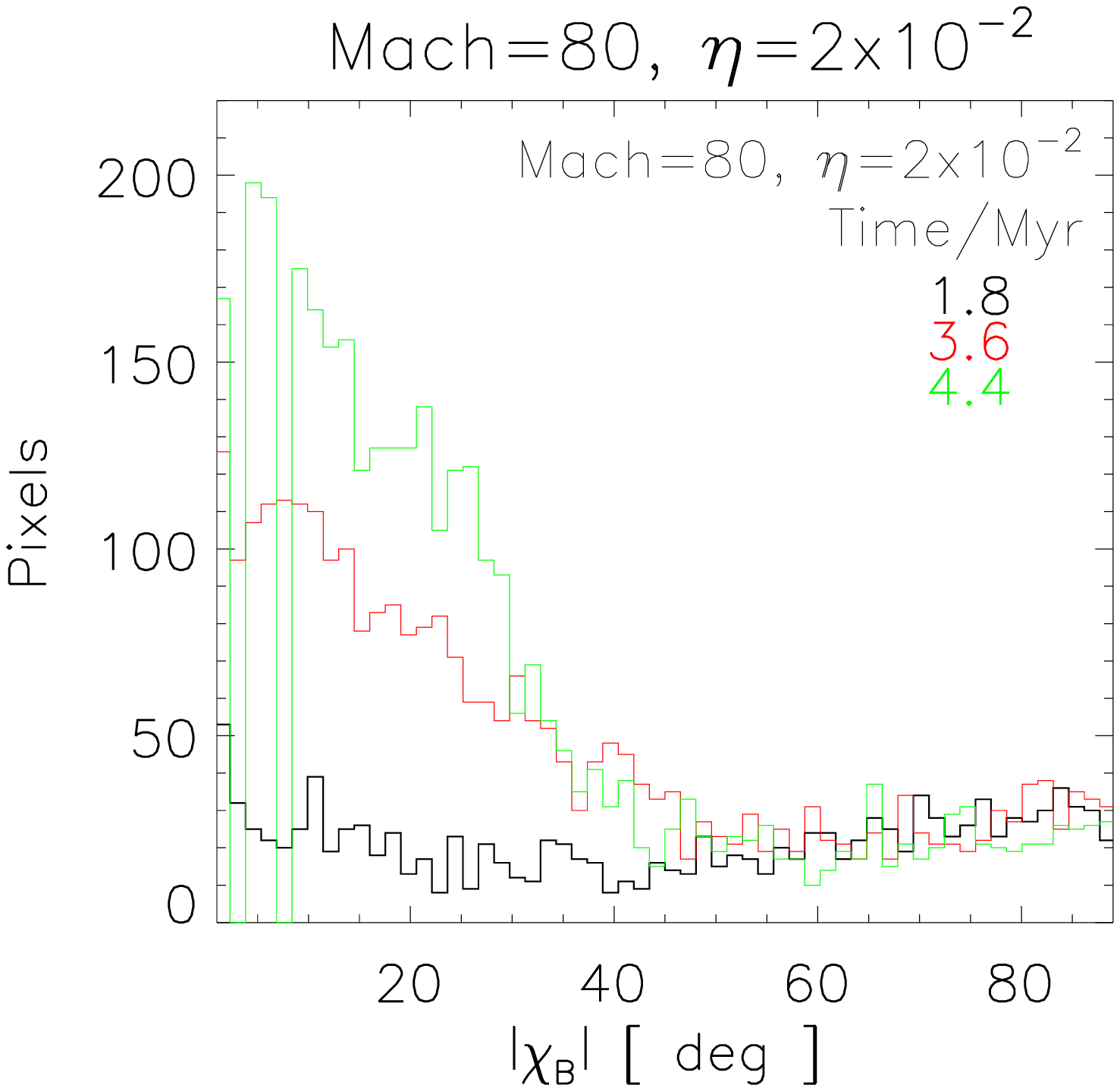}
     \end{center}
  \vspace*{-10pt}
   \caption
{Histograms of $|\chi_\mathrm{B}|$. 
%
	$\theta_\mathrm{v}$ increases from left to right, 
   column-wise.
%
}
   \label{histoArrayAngle}
   \end{figure*}

As the viewing angle increases 
we find 
the mean polarisation angle,
$\left< |\chi_\mathrm{B}| \right>$, 
decreases non-linearly 
(see Figure~\ref{histoArrayAngle}). 
%
   (panel \textit{b}, Figures~\ref{pol-sim01-90deg-a}--\ref{pol-sim04-90deg-a}). 
%
On average,
$|\chi_\mathrm{B}|$ diminishes for
about 9\de\ 
for viewing angles from 30~to 60~degrees, and 
about 4\de\ 
for viewing angles from 60~to 90~degrees. 
Cocoons 
have geometries that resemble 
prolate spheroids and thus 
\mfs\ inside them 
should relax 
easier %
along 
the jet axis than towards the equator.  However, to produce the
synthetic maps we follow two steps: 
(i) rotate the sources anti-clockwise, perpendicularly to the
jets, and (ii) project them onto the plane of the sky. Hence only 
the magnetic component along the jet 
axis (the horizontal one in the maps) is affected in this process
and grows in proportion to $\cos(\theta_\mathrm{v})$.

\begin{figure*}
     \begin{center}
~~~~~~~~~~~~~~~~~~~~~~~~~~~~~~~
{\large $\theta_\mathrm{v}=\,$30$^{\circ}$}
~~~~~~~~~~~~~~~~~~~~~~~~~~~~
{\large $\theta_\mathrm{v}=\,$60$^{\circ}$}
~~~~~~~~~~~~~~~~~~~~~~~~~~~~~
{\large $\theta_\mathrm{v}=\,$90$^{\circ}$}
~~~~~~~~~~~~~~~~~~~~~~~~~~ \\
\vskip.1cm
\includegraphics[width=0.302\textwidth,bb=40 75 470 395,clip=]
  {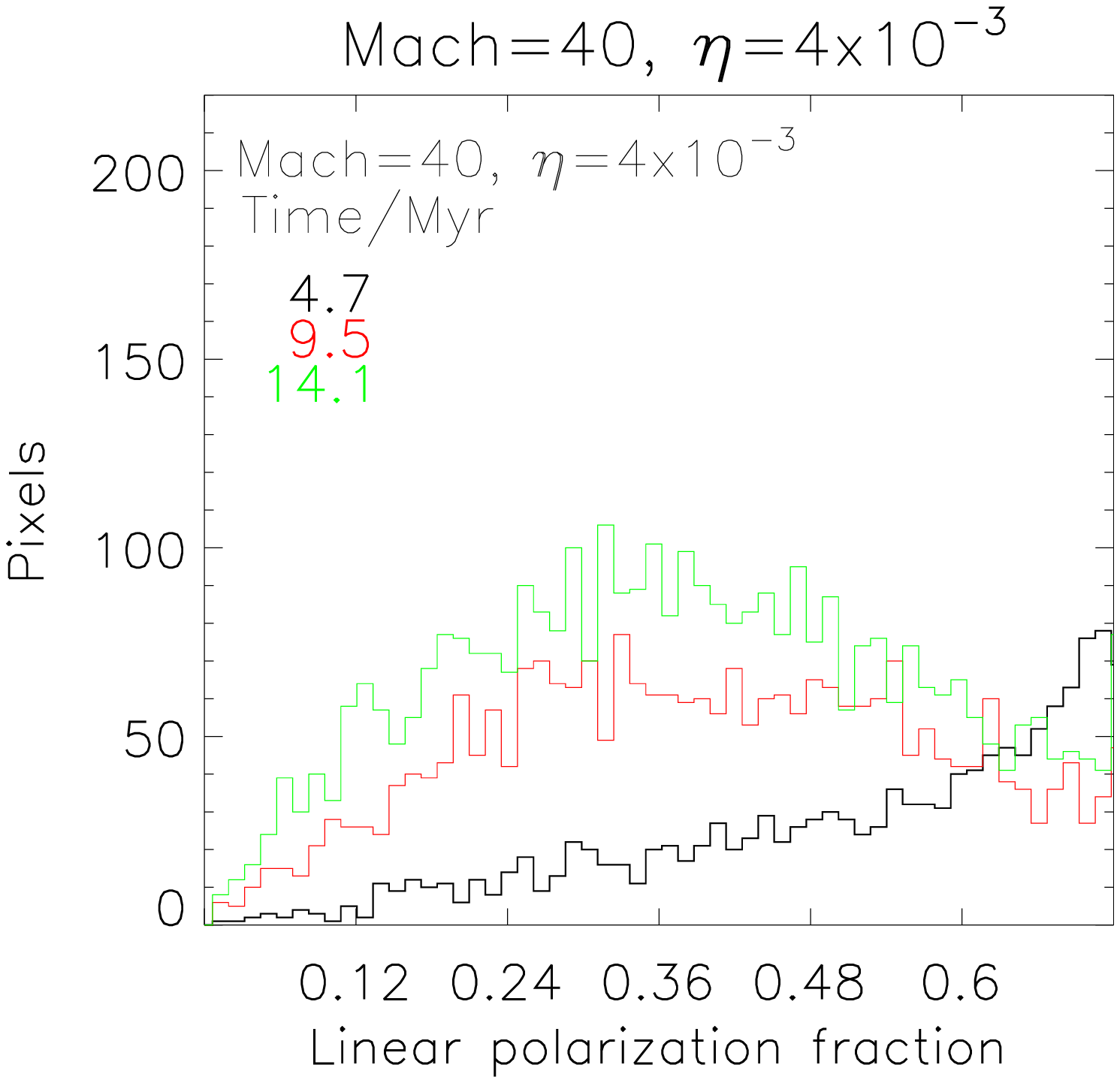}
\hskip-.12cm
\includegraphics[width=0.245\textwidth,bb=121 75 470 395,clip=]
  {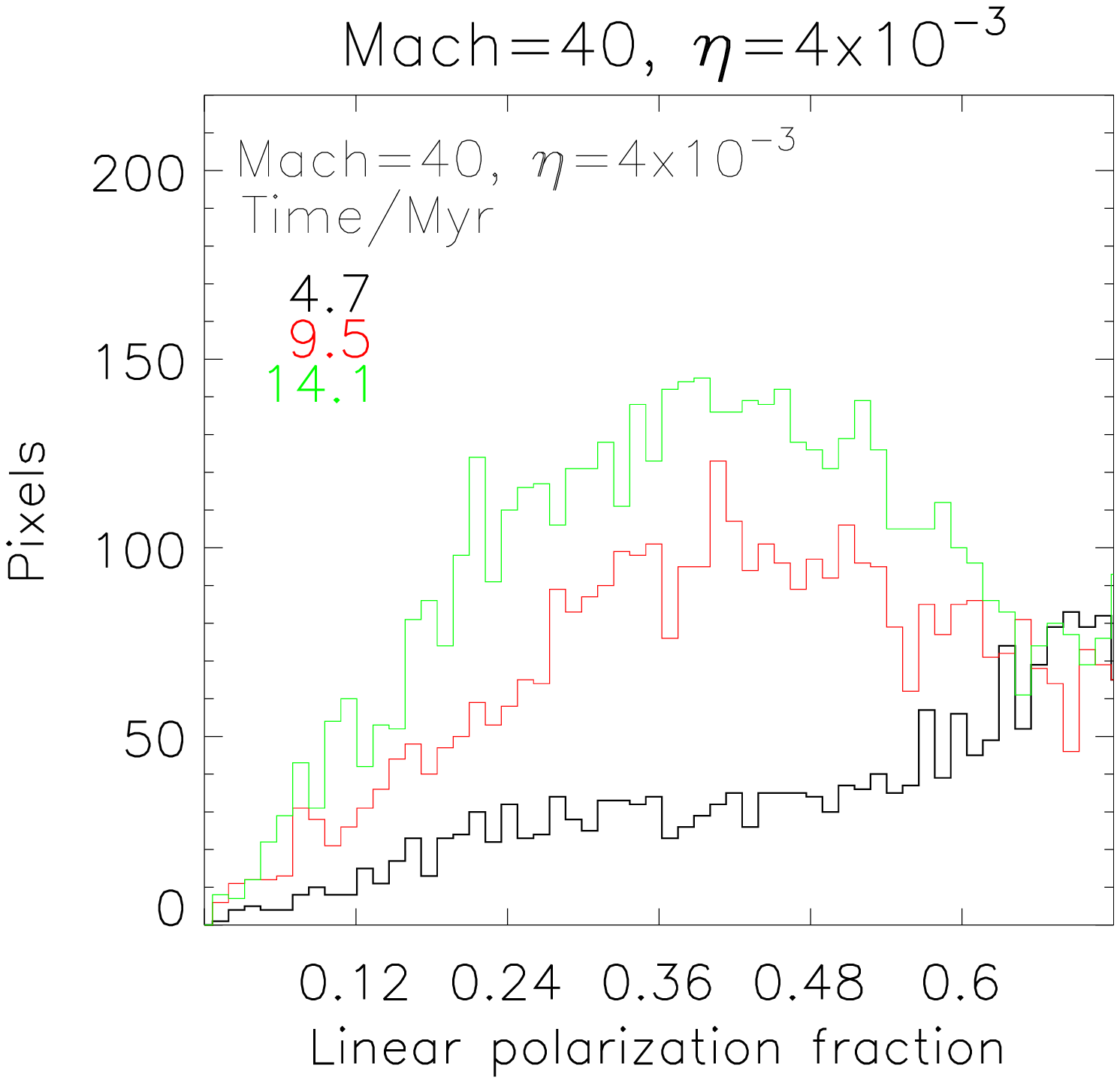}
\hskip-.12cm
\includegraphics[width=0.245\textwidth,bb=121 75 470 395,clip=]
  {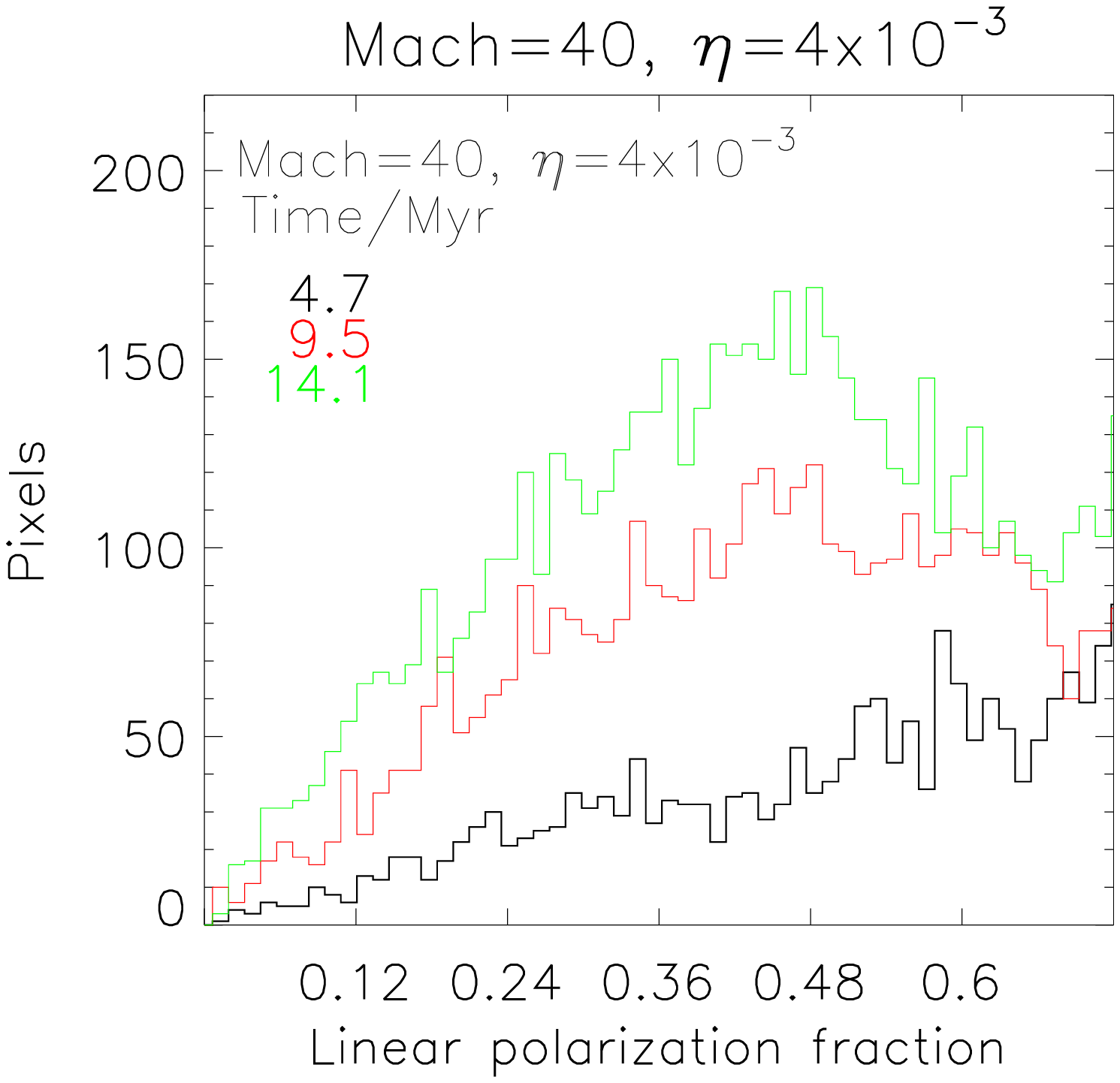}
%
\vskip-.062cm
\includegraphics[width=0.302\textwidth,bb=40 75 470 395,clip=]
  {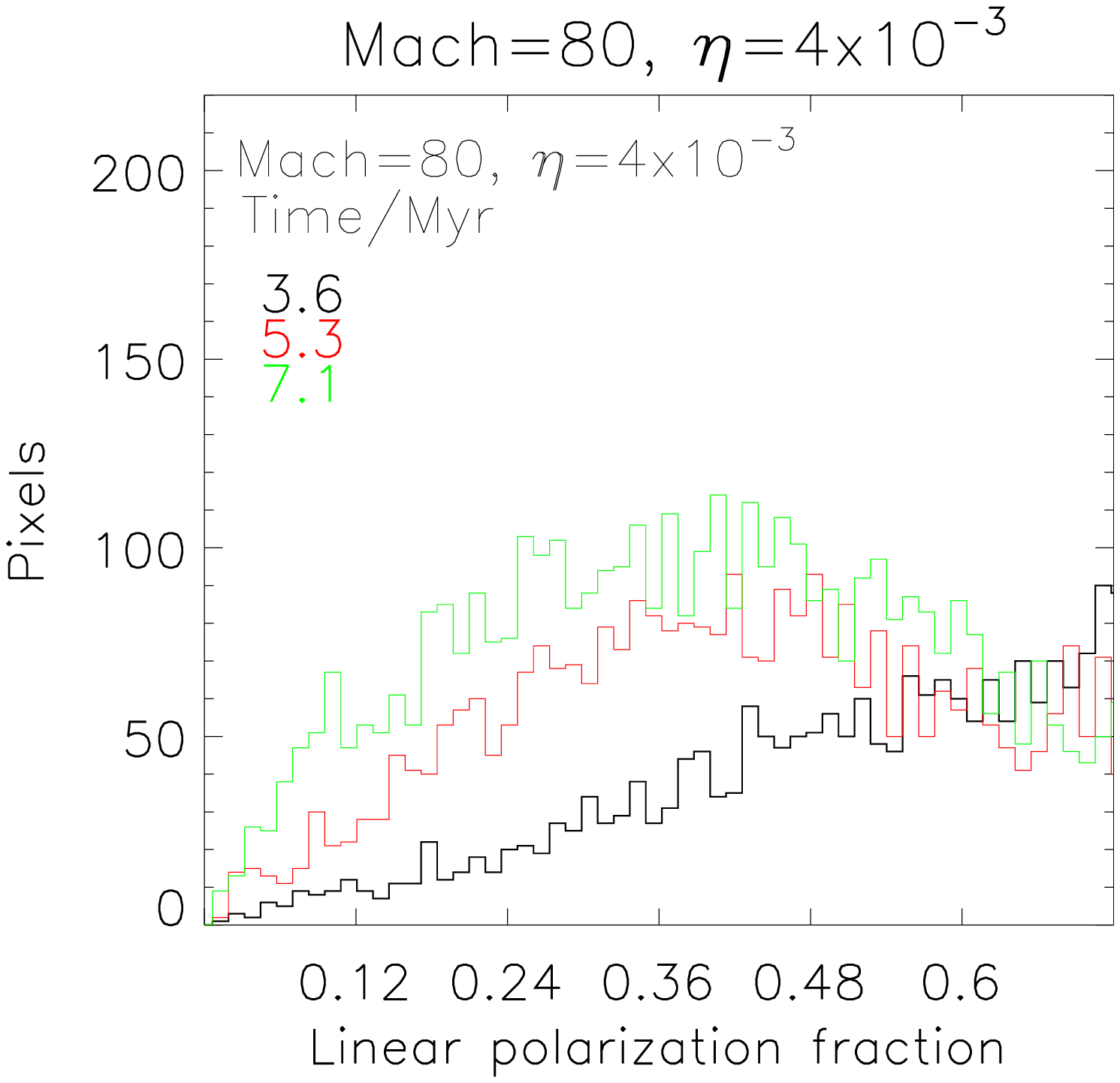}
\hskip-.12cm
\includegraphics[width=0.245\textwidth,bb=121 75 470 395,clip=]
  {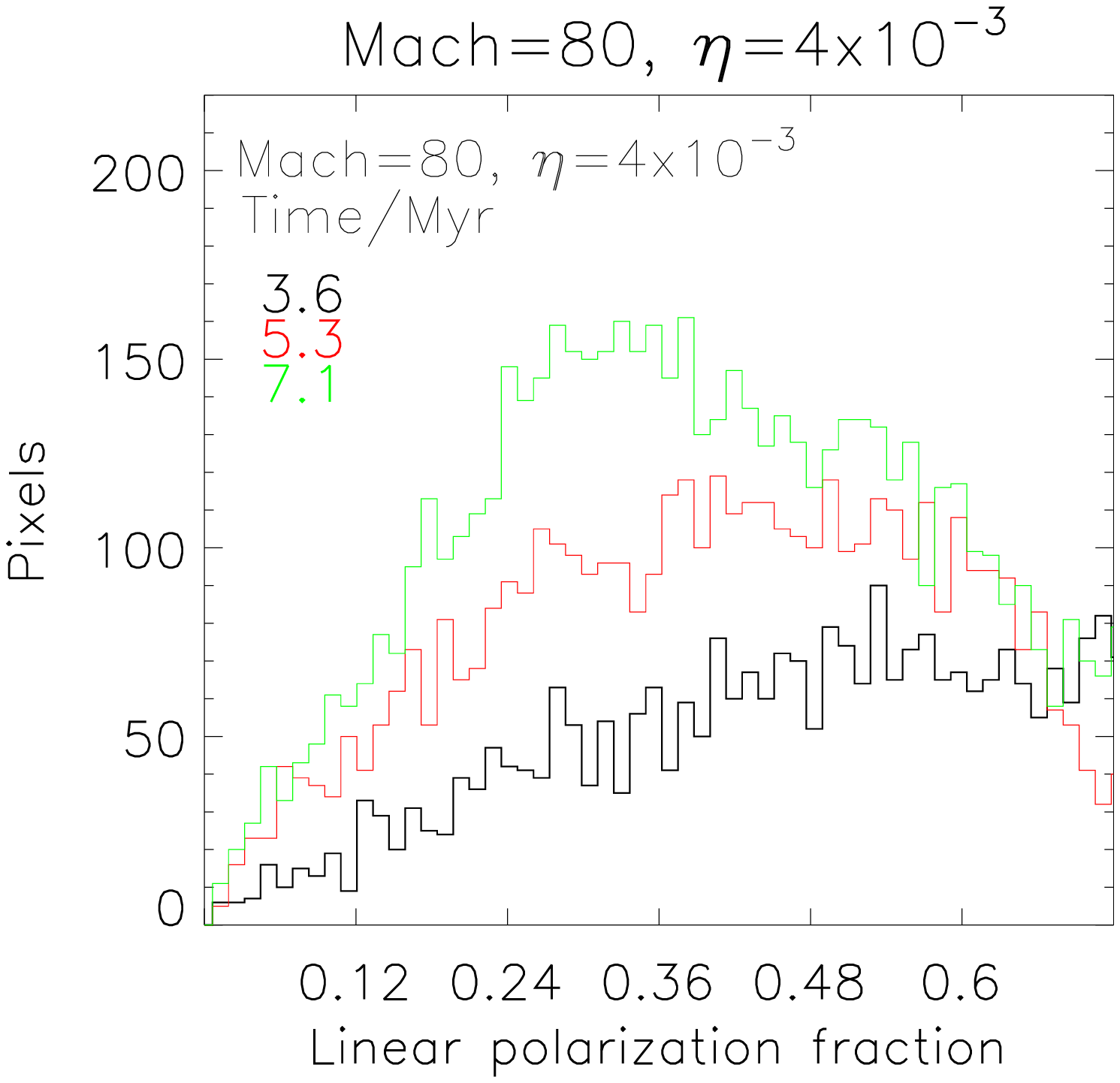}
\hskip-.12cm
\includegraphics[width=0.245\textwidth,bb=121 75 470 395,clip=]
  {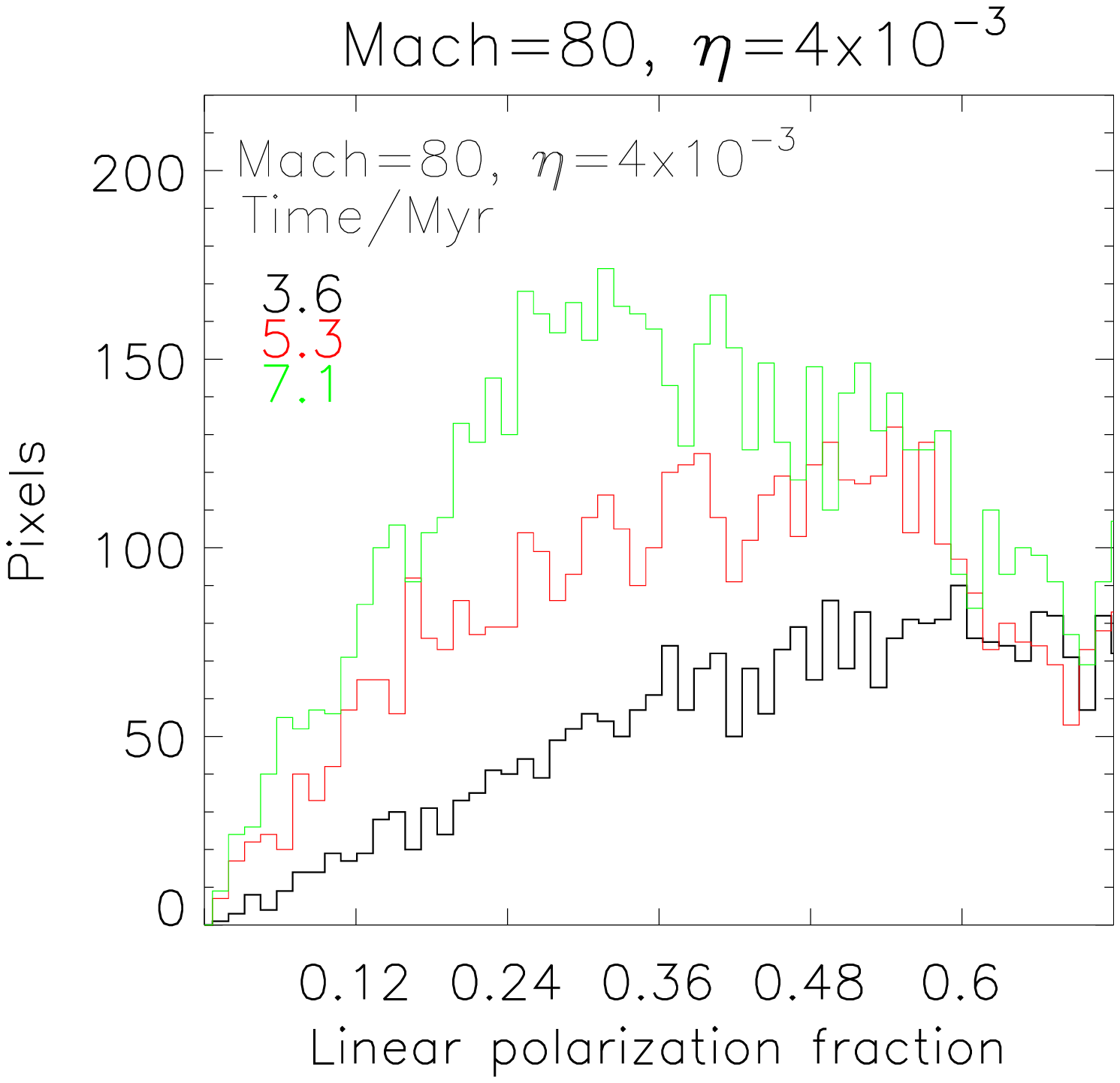} 
%
   \vskip-.062cm
   \includegraphics[width=0.302\textwidth,bb=40 75 470 395,clip=]
     {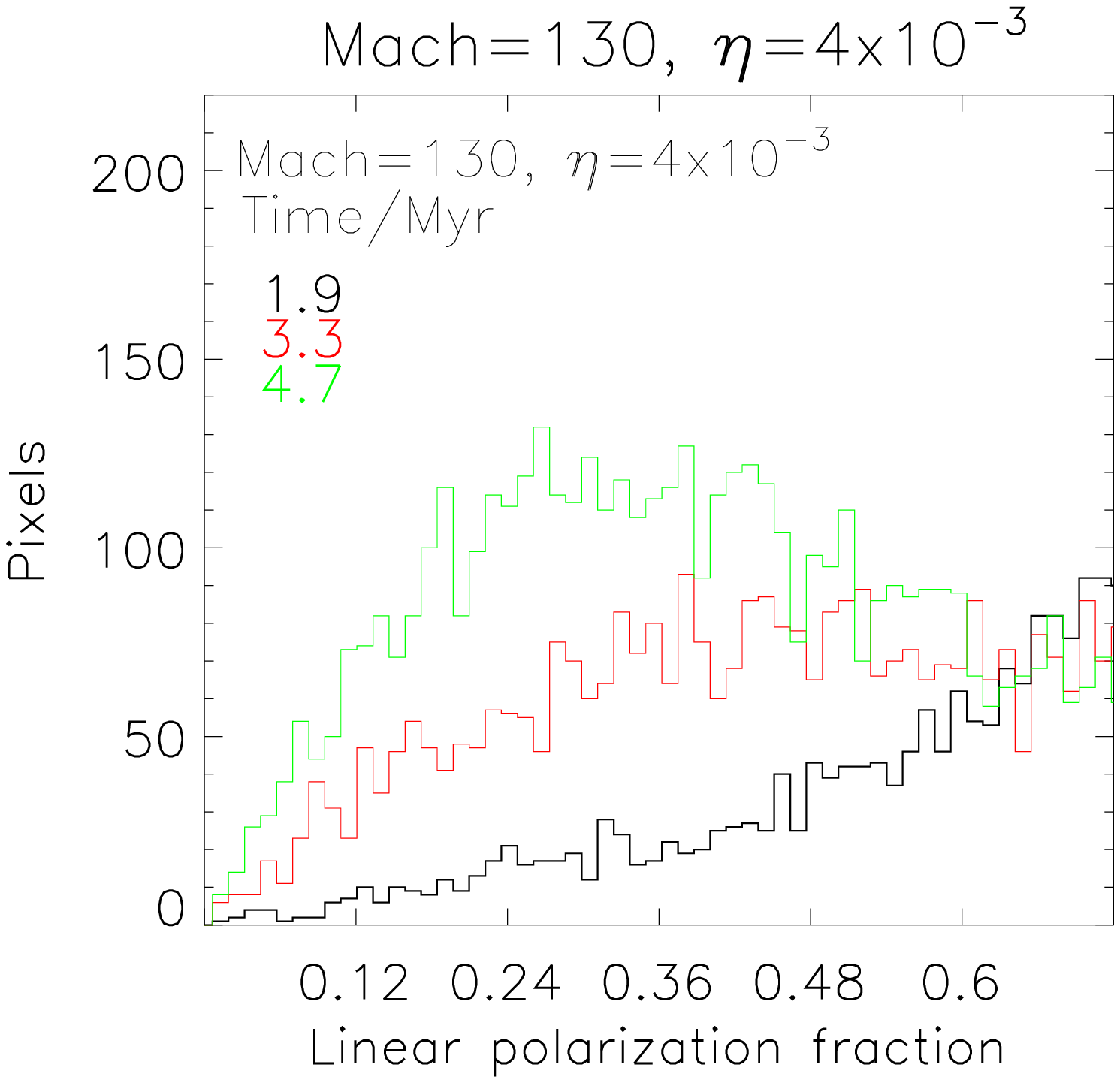}
   \hskip-.12cm
   \includegraphics[width=0.245\textwidth,bb=121 75 470 395,clip=]
     {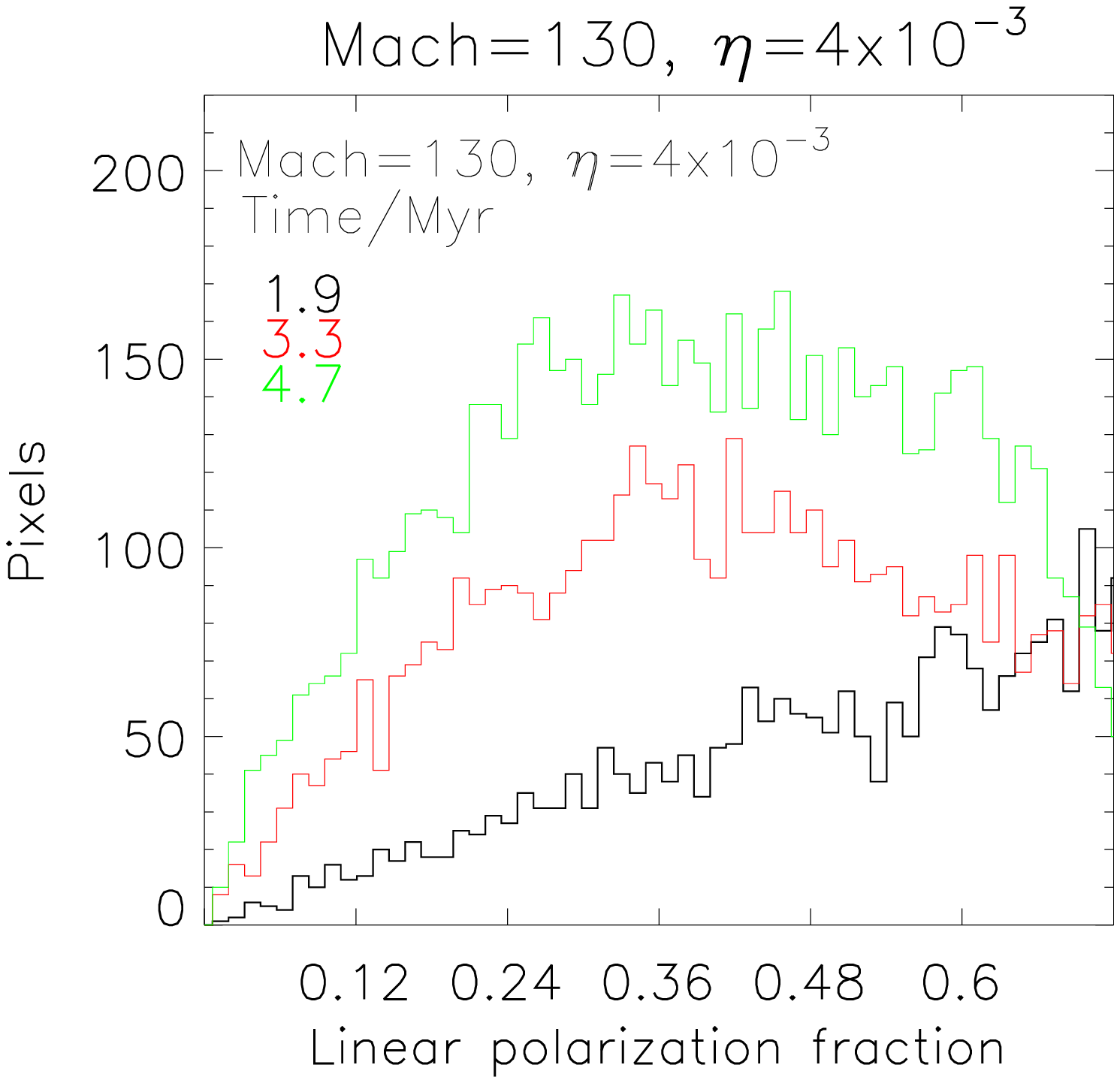}
   \hskip-.12cm
   \includegraphics[width=0.245\textwidth,bb=121 75 470 395,clip=]
     {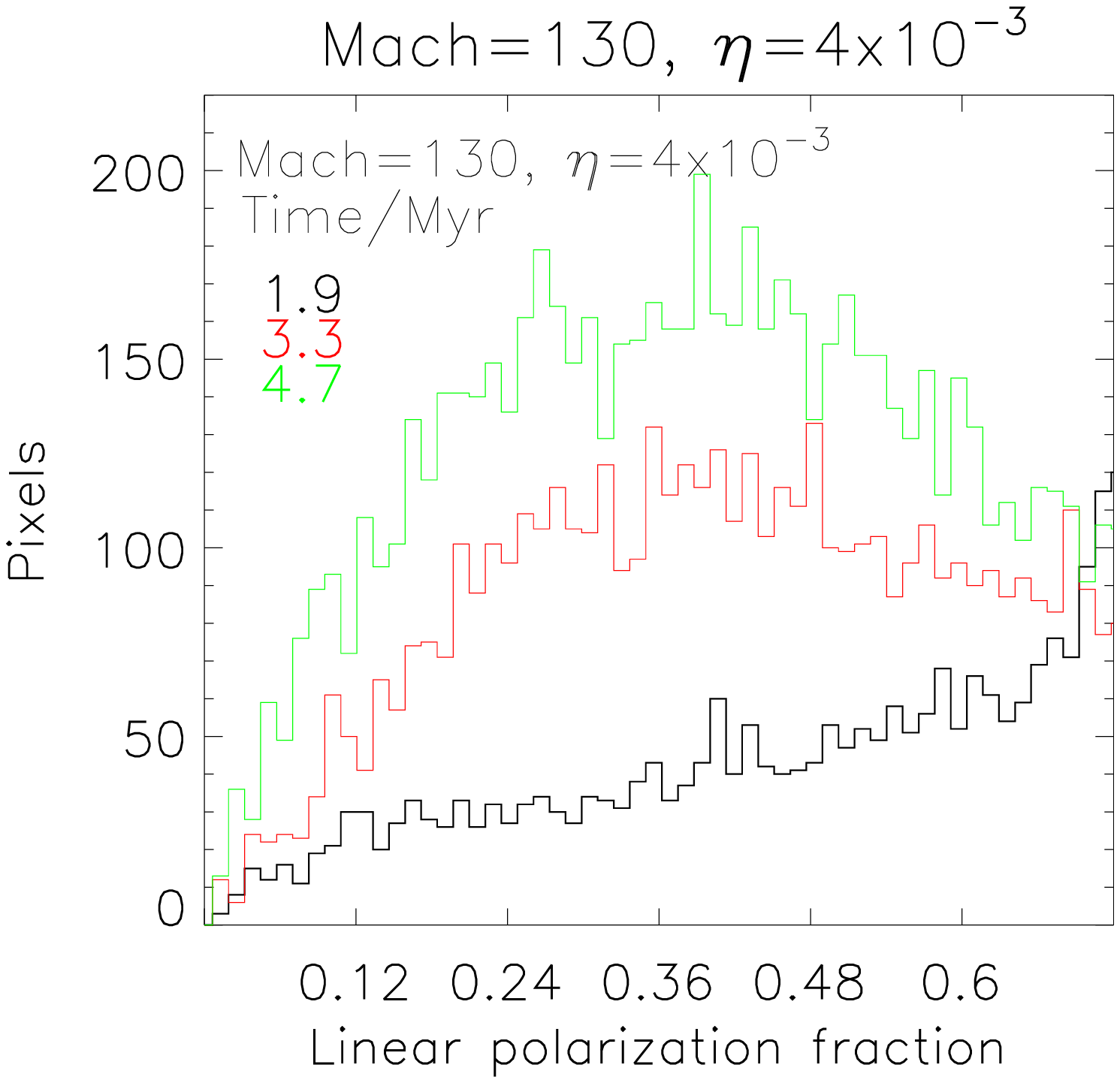} 
   %
\vskip-.062cm
\includegraphics[width=0.302\textwidth,bb=40 75 470 395,clip=]
  {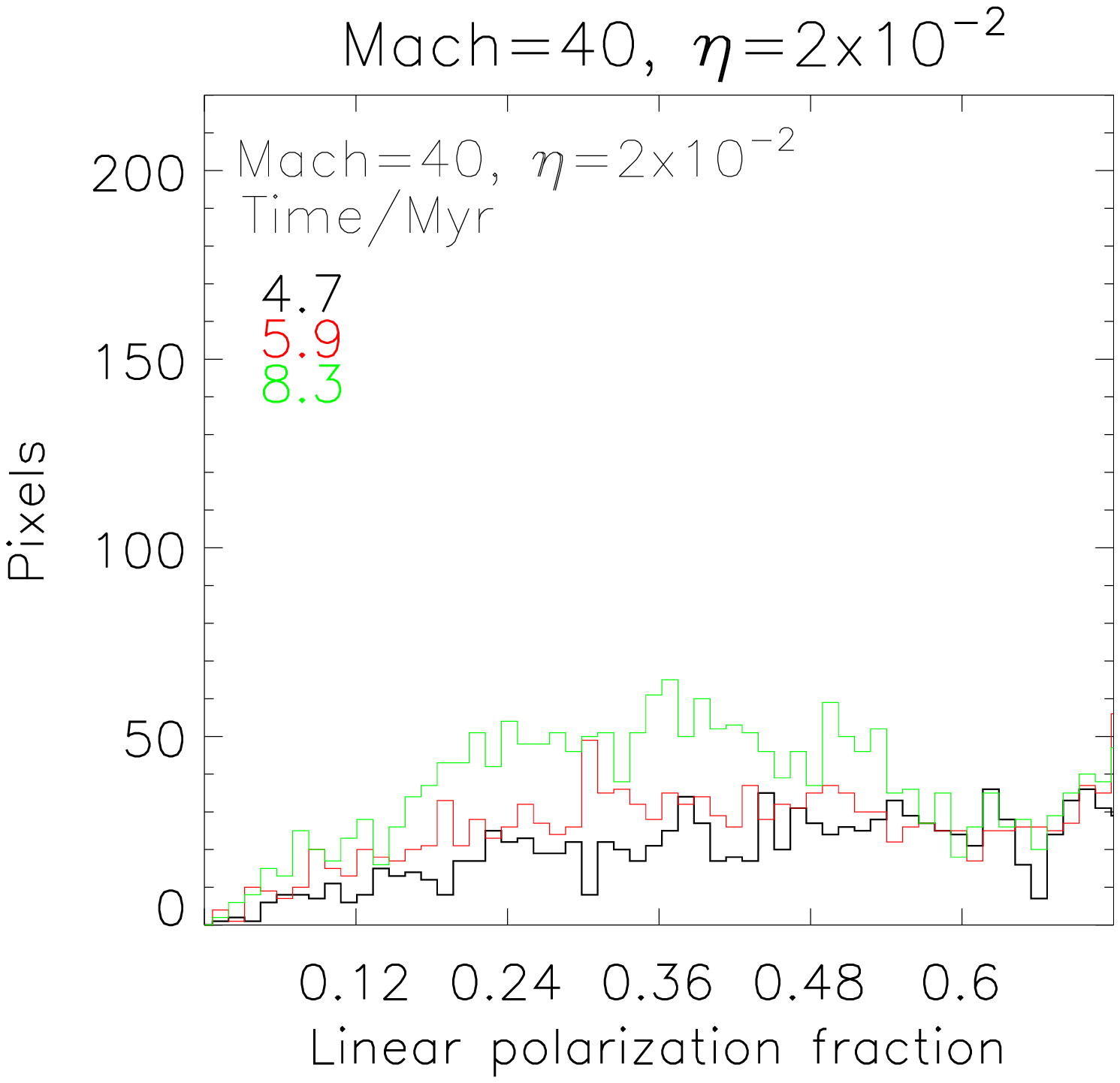}
\hskip-.12cm
\includegraphics[width=0.245\textwidth,bb=121 75 470 395,clip=]
  {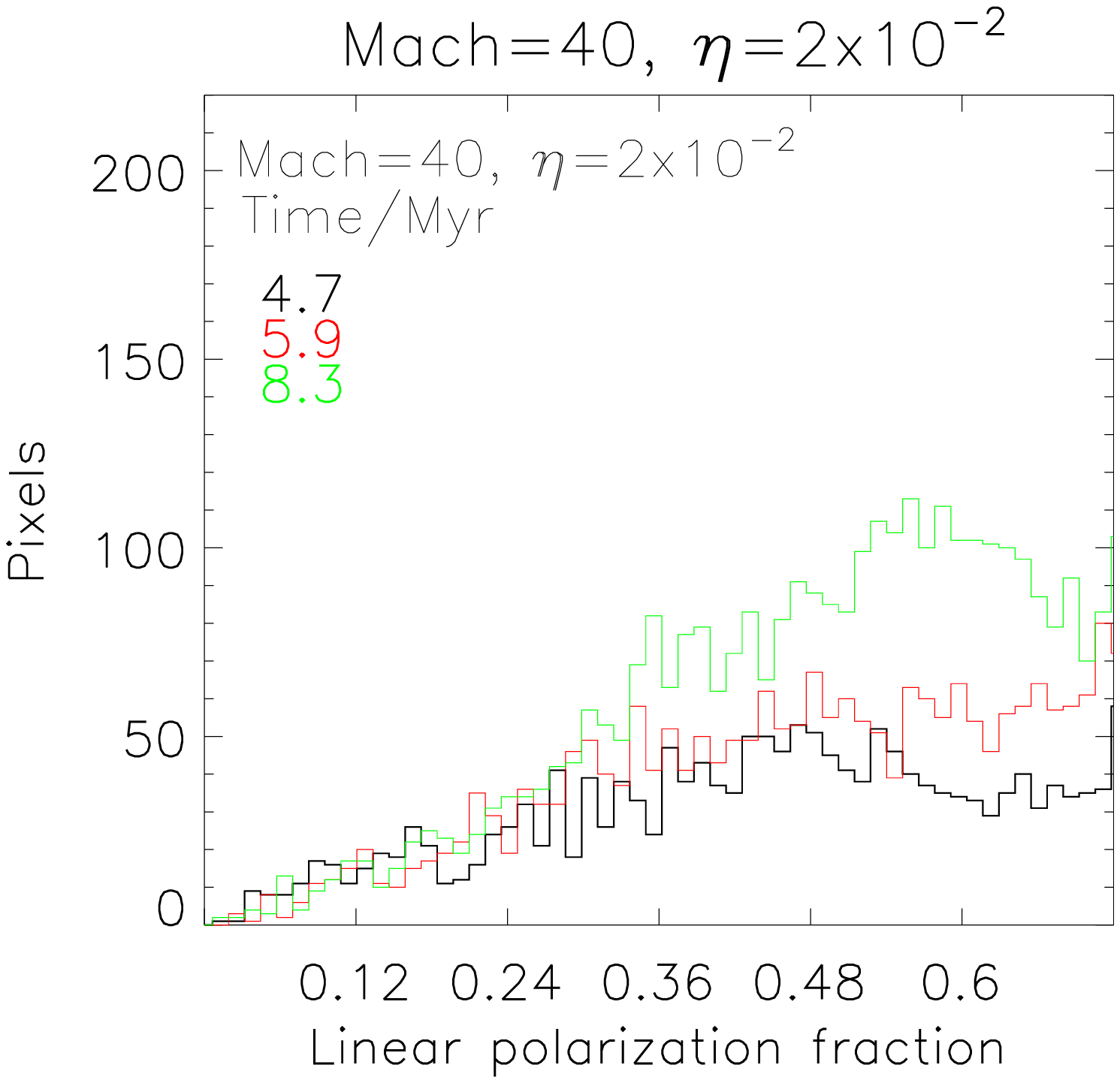}
\hskip-.12cm
\includegraphics[width=0.245\textwidth,bb=121 75 470 395,clip=]
  {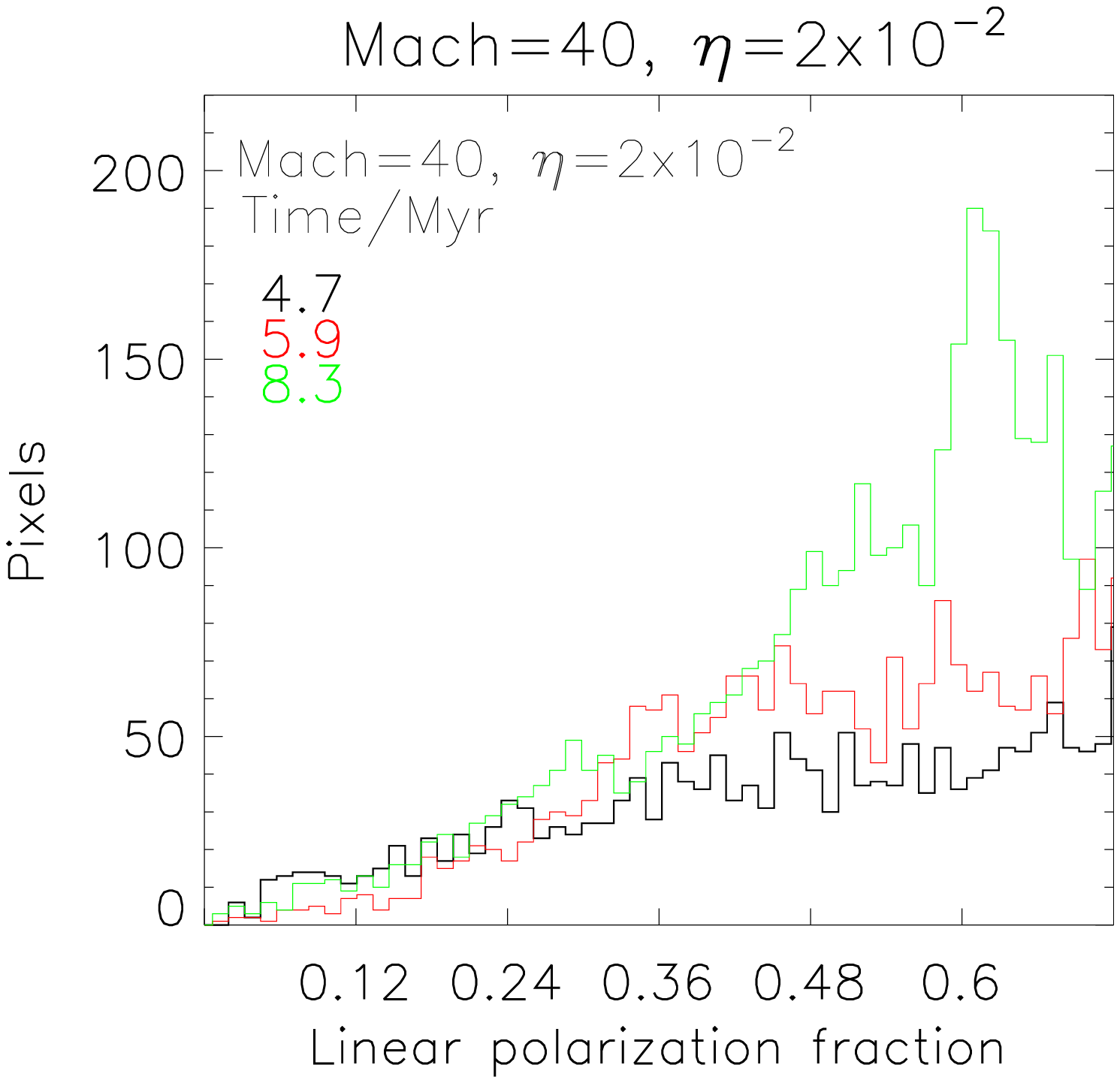}
%
\vskip-.062cm
\hskip.1cm
\includegraphics[width=0.302\textwidth,bb=40 10 470 395,clip=]
  {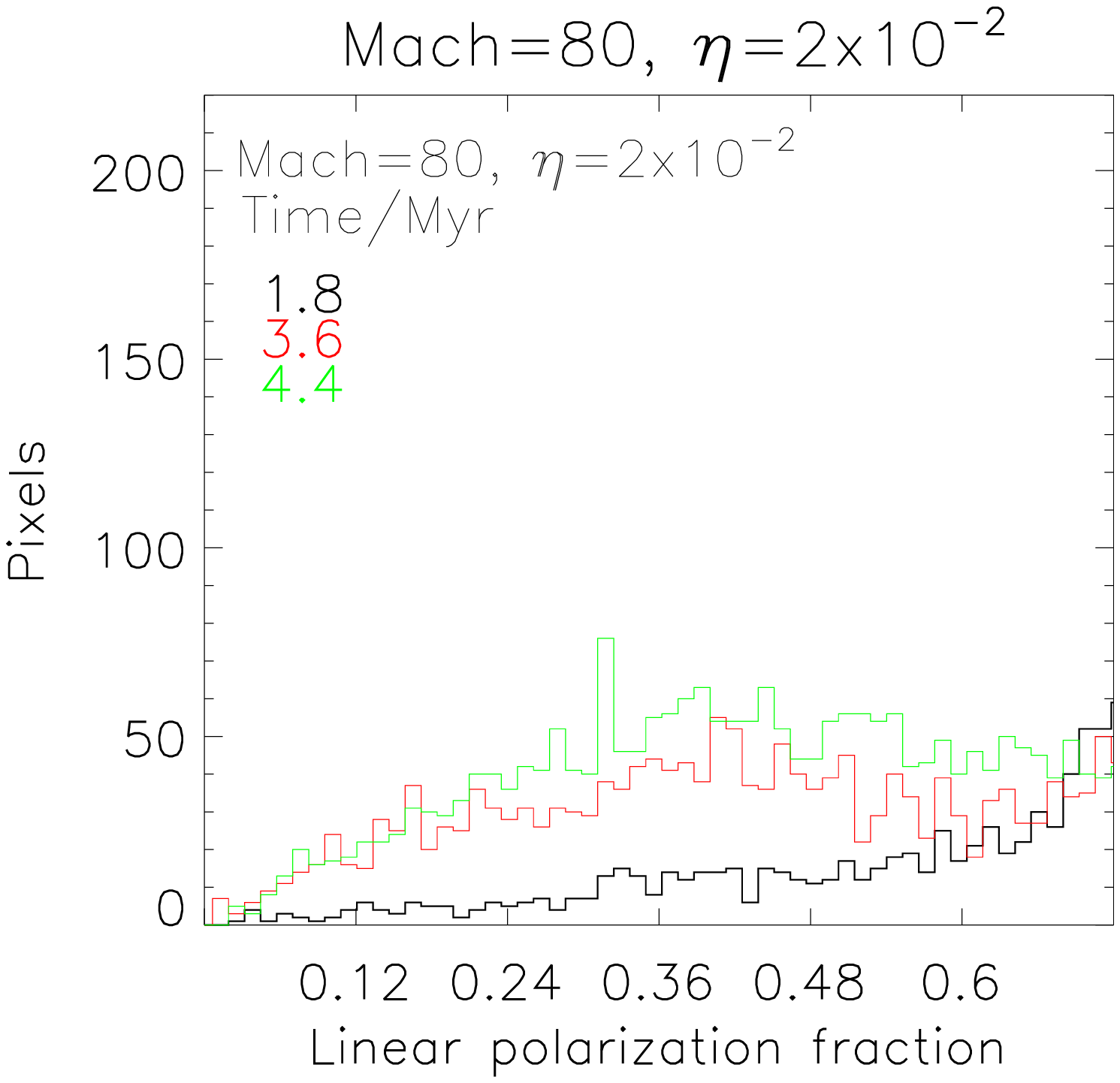}
\hskip-.12cm
\includegraphics[width=0.245\textwidth,bb=121 10 470 395,clip=]
  {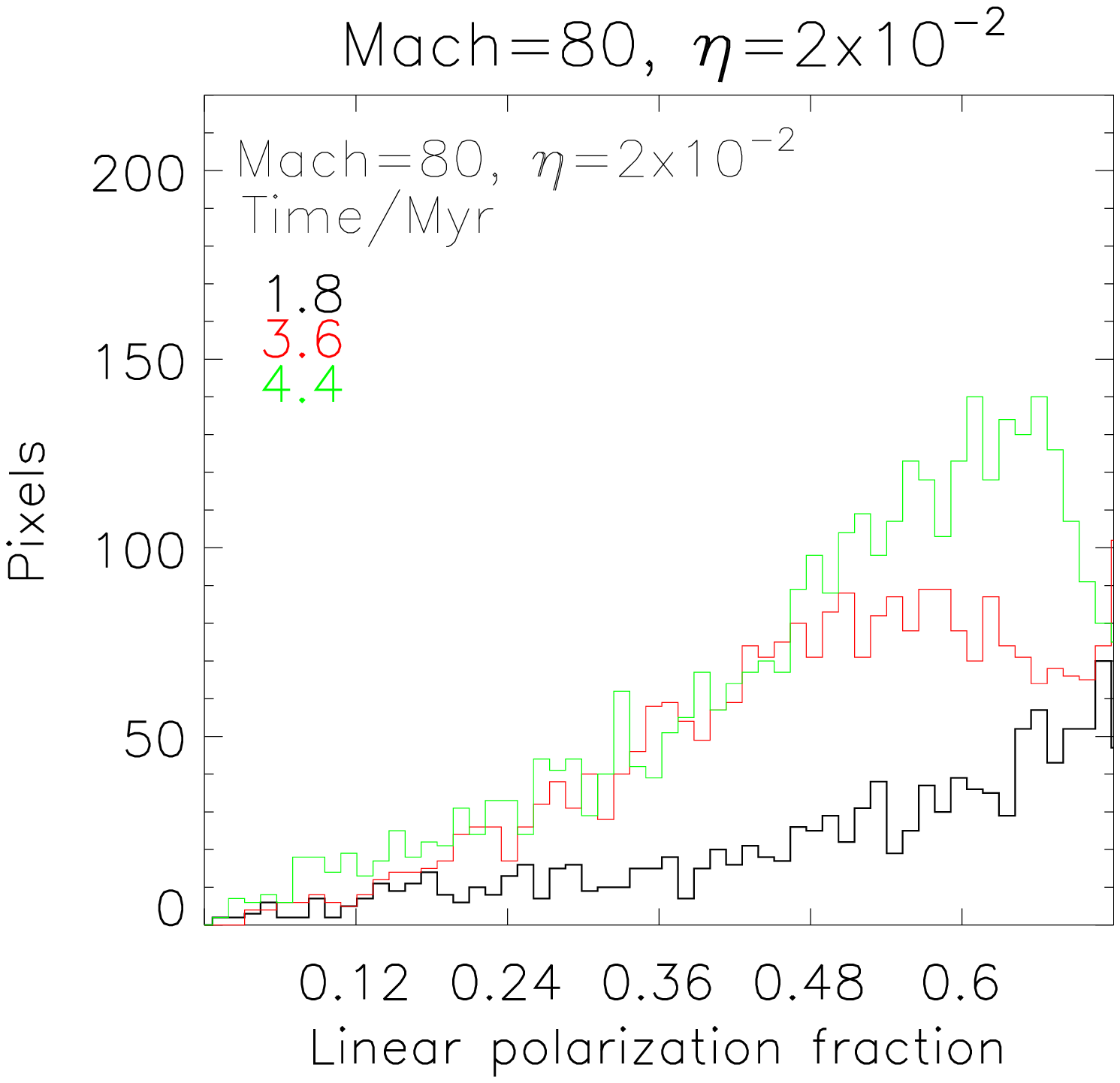}
\hskip-.12cm
\includegraphics[width=0.245\textwidth,bb=121 10 470 395,clip=]
  {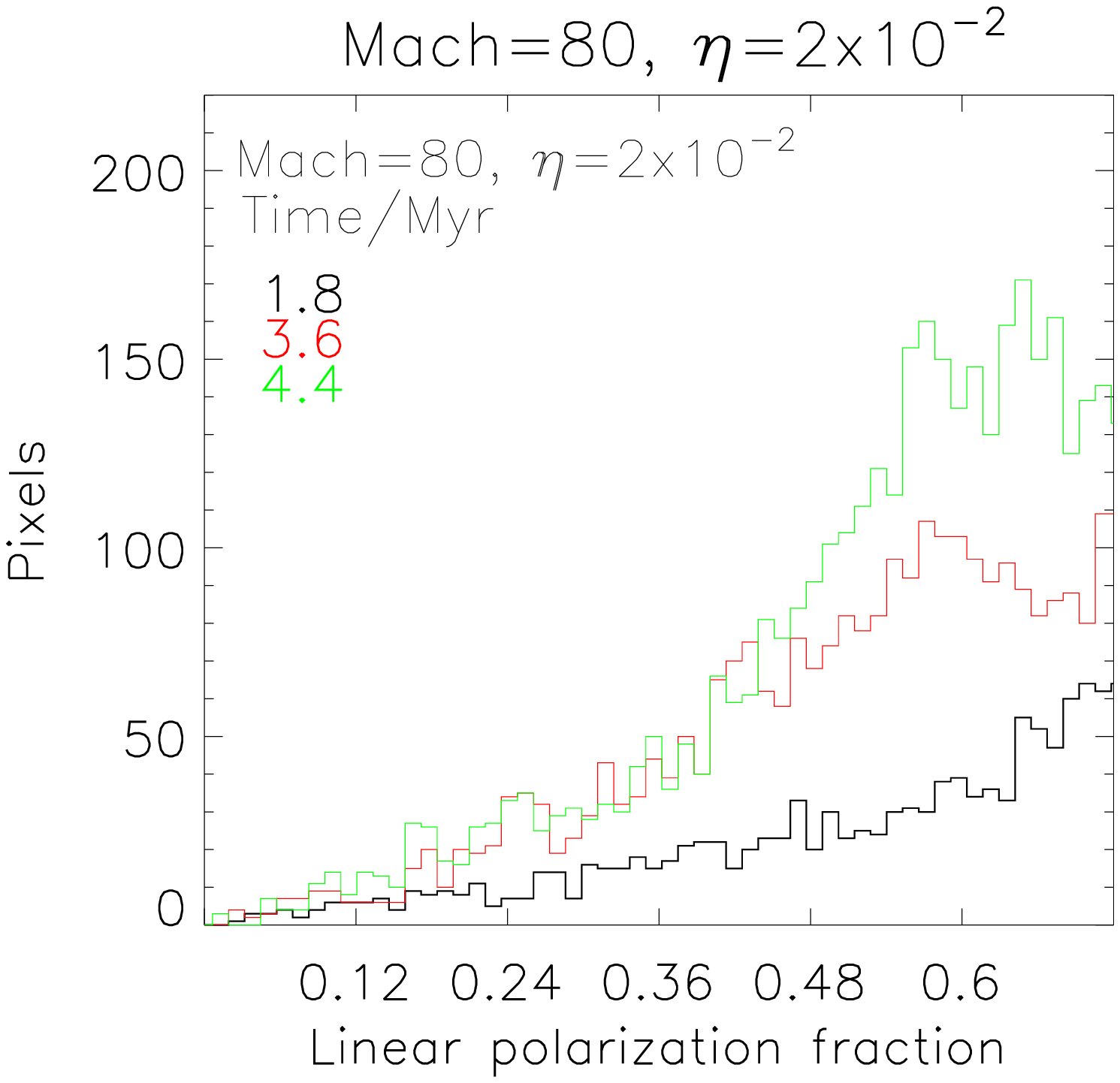}
  \vspace*{-10pt}
     \end{center}
   \caption{Histograms of the linear polarisation degree.
Panels are arranged as in Figure~\ref{histoArrayAngle}.
}
   \label{histoArrayDegree}
   \end{figure*}

The dependence of the degree of linear polarisation on the viewing
angle is relatively modest and 
particularly evident 
for $\eta=\,$0.02 
(Figure~\ref{histoArrayDegree}).
%
  %
  We see $\left< p \right>$ increases \hbox{about 7\%} from 
  $\theta_\mathrm{v}=\,$30\de\ to $\theta_\mathrm{v}=\,$60\de, 
  and also about $\sim\,$3\% from $\theta_\mathrm{v}=\,$60\de\
  to $\theta_\mathrm{v}=\,$90\de.
Hereafter, $\left< p \right>$ 
represents the arithmetic mean of $p$. 
For all $\eta$, the number of pixels in the
polarisation degree histograms consistently scales~up with the viewing 
angle, in relation to the projected area of 
cocoons.

\subsubsection{The role of the density contrast.}
\label{pol-vs-eta}

The jet-to-ambient density contrast is well known to be important
for the evolution of cavities 
formed by astrophysical jets (see
e.g. \citealp{vlj1}). Our synthetic maps show 
the density contrast also
plays an important role on the radio source polarimetry.  In general 
the projected area of 
sources is inversely proportional to $\eta$ 
in a non-linear way. 
Thus we see less \mfs\ 
in polarisation
measurements with \etaeq0.02 than, 0.004. 

Given a timestep and a viewing angle, we find 
the mean polarisation angle is typically $\sim\,$10\de\ 
smaller for 
\etaeq0.02 than for \etaeq0.004.  
We see the spatial distribution of the polarisation
angle is 
more uniform for $\eta=\,$0.02 
than for the lighter case. e.g. gradients greater than about 
10\de~per computational~cell 
(10\de$/\,$kpc) are less frequent in Figure~\ref{pol-sim01-90deg-a} 
(left column) than in Figure~\ref{pol-sim02-90deg-a}, 
corresponding to \etaeq0.02 and~0.004, 
respectively.

The statistical behaviour of the polarisation degree is very different.
Given a timestep and a viewing angle, we frequently find higher values of
the mean polarisation degree for \hbox{$\eta=\,$0.02} 
than for the lighter case. On average, 
%
   $\left< p \right>(\eta=\,$0.02)$\sim\,$47\,\%, 
   while $\left< p \right>(\eta=\,$0.004)$\sim\,$42\,\%. 
%
Moreover, the polarisation
degree histograms follow Gaussian-like distributions. 
The mean polarisation degree at large viewing angles increases with time
for the heavier jets, indicating that axial field line stretching gets
even more important with time. 
Conversely, 
it decreases with time for the lighter
jets, which shows that turbulence gets even more important with time
for the lighter jets. The polarisation degree of the
lighter jets does not depend on the viewing angle.

\subsubsection{Polarimetry evolution.}
\label{pol-vs-time}

The main features of the polarisation angle histograms seem to be
shaped during the early expansion phase of the model sources,
particularly for viewing angles of 60~and 90~degrees. Here, magnetic
fields tend to align with the jet axis \hbox{($\chi_\mathrm{B}=\,$0\de)}
as sources expand
%
   (panel \textit{a}, Figures~\ref{pol-sim01-90deg-a}--\ref{pol-sim04-90deg-a}). 
%
The considered histograms decline steeply up to
about~20\de to~40\de, and remain roughly constant for higher
$\chi_\mathrm{B}$.  The constant part is at a very similar level
for all viewing angles of a given simulation. These findings
correspond to the effects of isotropic turbulence, in combination
with the stretching of field lines in cocoons, predominantly along
the jet direction.

The fractional polarisation evolves quite differently. At the
first timestep (Figure~\ref{histoArrayDegree}, black profiles), we see that the
$p$ histograms are fairly similar and show a linear relation, rising
monotonically towards larger $p$.  
As the cocoons with $\eta =\,$0.004 
grow, the profiles evolve into a
peaked distribution with a broad peak between about $p=0.2$ and,~0.5.
In contrast, the profiles
for ($\eta =\,$0.02, 
$\theta_\mathrm{v} \geq \,$60\de) always peak at $p>0.5$.
%
  The fractional polarisation tells us about two things: (1) the
  degree of alignment of the field vectors that contribute to a
  given line of sight, and (2) the number of pixels long that line
  (assuming their contribution is different from each other). 
  Hence we see the polarisation generally decreasing for $\theta=\,$30\de.
  In cocoons where $\eta =\,$0.004 
  we see a stronger $p$ decline at early times than 
     later on. 
This occurs because their
  expansion slows down at late times, as these sources approach pressure
  equilibrium with the ambient medium.  In cocoons where $\eta=\,$0.02, 
  on the other hand,
  we see a slow sideways growth and thus $p$ drops very slowly.
  Moreover, we see higher polarisation for larger $\eta$, again
  reflecting that high density jets have more ordered magnetic
  fields and blow thinner cocoons.
%

%
%

We note that an additional set of synthetic polarisation maps (not
shown) was produced assuming a spatially uniform distribution of
synchrotron emitting electrons $[$i.e. 
$p_c({\bf x},t)=\,$1 in (\ref{stokes}), for all ${\bf x}$ and $t]$. The polarimetric
distribution of such maps was found to be very similar to the ones 
discussed in this section of the paper. Our results are not, therefore, 
sensitive to details of the electron distribution.

Radio source polarimetry is related with the study of cluster magnetic
fields because they induce Faraday rotation and depolarisation
on the radio source emission \citep{pacholczyk63,burn66}.  
Faraday rotation maps contain information about the ICM's
magnetic structure (for a review see \citealp{carilli02}). In a
sequel paper (Huarte-Espinosa, Krause, \& Alexander 2011b, in prep.)
we will investigate the evolution of cluster magnetic fields using
statistical analysis on synthetic RM observations which are produced
using the 
expanding 
model sources we present here.


\section{Discussion} \label{discu}

About a handful of studies on synthetic synchrotron emission and 
polarimetry of extragalactic radio sources exist in the literature. 
We are not aware of any study that uses magnetic fields evolved in a
magnetohydrodynamic simulation self-consistently with the jet, as
presented here.

\citet{jones88} modeled relativistic jets with a turbulent magnetic
field ansatz, advected with the flow velocity of the jet, to study
the relation between linear and circular polarisation in compact
radio sources. The underlying hydrodynamic simulation is a conically
expanding beam. With this ansatz, he gets a few, up to 23~per cent,
linear fractional polarisation. Though we start from a similarly turbulent
field in our initial injection region, we get about 50~per cent linear
fractional polarisation in our beams. The reason is the order induced
by the field line stretching described in more detail in
Section~\ref{flow}. Models of jet collimation and acceleration
typically require a poloidal field near the source \citep[e.g.][and
references therein]{PF10}. The coherence length of this initial field
should be small compared to observed jet sizes. Hence, stretching  
of the magnetic field in the beam seems to be an unavoidable
consequence. The effect is also found by Gaibler et al. (2009). The latter
study is however the only one we are aware of that 
has 
employed a zero gradient boundary condition in the jet nozzle. 

Axisymmetric hydrodynamical simulations of collimated light jets,
similar to our approach, were employed by
\citet{matthews90a} to simulate the advection and deformation of 
passive \mfs\ set up with an initial isotropic random geometry, similar
to \citet{jones88}.
Matthews \& Scheuer implemented \mfs\ using passive 
tracer particles and followed the distortion of the fluid, at the
respective position of tracer particles, by the velocity field
computed in the hydrodynamic simulation. This gives a reasonable
approximation of the magnetic field structure for dynamically passive
magnetic fields. We also have a dynamically passive field, with a very
similar initial, and slightly different nozzle boundary condition. In
contrast to them, we do a full MHD treatment for the magnetic
field. We confirm almost all of their results regarding the magnetic
field structure: Matthews \& Scheuer discuss in detail the toroidal
and the poloidal stretching mechanism. As argued above, we believe 
the toroidal
stretching mechanism is mainly responsible for the magnetic energy
increase 
in 
cocoons. We do not observe a dominant toroidal field
directly because non-axisymmetric shear converts this component to
poloidal field. We generally see a predominantly axial field component
in cocoons, consistent with their poloidal stretching mechanism.
In contrast to them, we also find axially stretched and amplified magnetic
fields in the beams. 

As \citet{matthews90a} do, we find field line stretching along the
contact surface that separates cocoons from the ambient medium. In
a resolution study they show that the extent of that region gets
smaller at higher resolution, but the field strength increases due
to the increased shear.  They also address synchrotron losses in
the energy distribution of relativistic electrons. Due to such
losses, they find that the aforementioned shear layer is very weak
in synthetic radio maps.  In our maps these features appear as edge
enhancements and are likely a numerical artifact because our treatment
does follow synchrotron losses.  In reality, the two fluids may
slip easily and the shear layer may be insignificant. This depends
on the magnetic viscosity of the plasma and is beyond the scope of
this discussion.

Moreover, as we do, \citet{matthews90a} find filaments in the
synthetic emission
images, but they report close to maximum fractional polarisation. We find lower, more
realistic, fractional polarisation values in our simulations,
especially for the lighter jets. The main
reason for this difference is the breaking of axisymmetry: This allows
for 3D turbulence in cocoons and for different directions of 
magnetic field vectors along the azimuthal direction.
Yet, we also see fractional polarisation values in the cocoon body, 
far away from the beams and the edges, which are still somewhat 
high. This might
be a resolution issue: the magnetic field energy spectrum is close to
Kolmogorov, which we have checked for the final snapshots of all our
runs. Therefore, dominant structures are the large scale ones, which
we should be able to capture. However, the roughly 50 cells we have over
the fatter cocoons might still be too little to capture some important
small scale structure that could reduce the fractional
polarisation. Our simulations show the fractional polarisation is
very similar for different jet velocities. Also, as
\citet{matthews90a}  have already noted, cocoon magnetic fields are
largely independent of the initial conditions prescribed at the base
of the beams. It is therefore unlikely that there is something
fundamental to the cocoon structure that we miss.  Another reason for
low fractional polarisation might be that our jet densities may still
be too high. We find the cocoon width, which is mainly regulated by the
jet density, is an important factor for the polarimetry. Observed
cocoons are usually wider relative to the beam than the ones we 
produce here. This fact indicates lower jet densities in the observed radio sources 
\citep{alexander96,vlj1}. Hence, low fractional polarisation might be
yet another consequence of jets being very light compared to their
surroundings. 
%
   Finally, \citet{matthews90a} found small regions where field
	amplification and therefore synchrotron cooling became very
	significant in their simulations. In our 3DMHD simulations
	we see filaments in the cocoons (Figures~4-8, right column)
	and that magnetic fields there are about 
	an order of magnitude stronger than the mean field.
   Thus we confirm the findings of \citet{matthews90a}.
%


\citet{treguillis01} carried out 3D-MHD simulations of a
jet with $\eta=\,$0.01, Mach=80 and
a helical magnetic field, the axial part of which extended throughout
the computational domain.
These authors studied the diffusive shock acceleration and transport of synchrotron 
relativistic electrons. We do not follow such processes. Then,
\citet{treguillis04} produced detailed synthetic observations of both the
synchrotron and the X-ray --\,due Compton scatter from CMB photons.
They emphasise that along the lines of sight that pass through strong
shocks, most of the emission may come from regions close to the
shock, and thus have close to maximum fractional polarisation values. We
might miss some of such regions due to the limitations of our simple
model for the distribution of relativistic electrons. However, the
emission from the bulk of 
cocoons cannot be dominated by such
features, as the fractional polarisation we predict for such regions
is too high (compare above). This would mean that real radio lobes are
relatively uniformly illuminated by relativistic electrons and are
not dominated by relatively few isolated shock features. 


\section{Summary and conclusions}
\label{conclu}

We carried out 3D-MHD numerical simulations and synthetic observations
to model magnetic fields in expanding FR~II sources located at
the core of a non-cool core galaxy cluster. A stratified fully ionized
ICM was implemented, threaded by randomly tangled \mfs\ with a
Kolmogorov power spectrum.  Collimated, hypersonic
and bipolar jets were injected
in the centre of the computational domain.  The geometry of the
jets' \mfs\ is initially random, and then shaped by the dynamics
of jets. Jets form cocoons filled
with light gas and magnetic fields, the structure of which is
determined by both the jets' backflow, via shear and compression,
and the cocoon expansion.

We have presented five simulations exploring the parameter space given
by jet-to-ambient density contrasts of 
$\eta=$\{0.004,~0.02\}, 
and jet velocities of $v_\mathrm{j}=$\{40,~80,~130\}\,Mach.  We
use the resulting model sources to produce synthetic synchrotron
emission and linear polarisation maps at viewing angles of
$\theta_\mathrm{v}=$\{30\de,~60\de,~90\de\}.  
The simulations have taught us the following.

While we do not inject magnetic energy at the jet nozzle, the
magnetic energy in jets, and their host cocoons, increases with
time. The amplification is stronger for wider cocoons, which are
obtained for lighter and faster jets. The main amplification mechanism
is the toroidal field line stretching (\citealp{matthews90b}; 
Gaibler et~al. 2009).
The 
toroidal field is however quickly converted to poloidal field
and the resulting field structure is hence a competition between
MHD-turbulence and poloidal field stretching. Lighter jets are more
turbulent and their magnetic field is therefore less aligned with
the jet axis.

Our synthetic polarisation maps are in good agreement with radio
observations \citep[e.g.][]{johnson95,gilbert04,mullin06}. We
generally see B-vectors that are parallel to the jet axis, tangent
to the source boundaries and perpendicular to strong emission
gradients. The degree of linear polarisation along both the jet
axis and the source boundaries is higher than both inside and between
radio lobes.

The cocoon magnetic structure shows a strong relation with $\eta$
and a rather weak relation with $v_\mathrm{j}$. 
In our polarisation maps this occurs
because the projected sources' area onto the plane of the sky is
proportional to the cocoons' volume. The intrinsic polarisation angle
distribution is consistently more uniform for \etaeq0.02 
than in the lighter case. The mean polarisation angle is $\sim\,$10\de\
smaller when \etaeq0.02 
than in the lighter case.  Also, the intrinsic linear
polarisation degree in the \etaeq0.02 
case is higher than in lighter
sources. i.e. when \etaeq0.02 
we see $p$ within 46-51~per cent in the
cocoons and, \hbox{$\sim\,$63~per cent} at the sources' edges. Conversely, when
\etaeq0.004 
we see $p$ within 25-45~per cent in the cocoons 
\hbox{and, $\sim\,$63~per cent} at the edges. 
   Even for our lighter cocoons, the fractional polarisation is somewhat
   high away from the edges and beams, 
which might be a resolution issue or due to the fact that our
cocoons are thinner than 
those 
of most observed FR~II radio
sources, which is related to the jet density.

The distribution of the polarisation angle (magnetic vectors)
depends on the viewing
angle between jets and the line of sight, $\theta_\mathrm{v}$.  On
average we see $\left< |\chi_\mathrm{B}| \right>$ decreases about
9~degrees as $\theta_\mathrm{v}$ goes from~30\de\ to~60\de,
and about~4~degrees as $\theta_\mathrm{v}$ goes from~60\de\ to~90\de. 
In contrast, only $\left< p \right.$ (\etaeq0.02) 
$\left. \right>$ shows an increase of about~7\% as $\theta_\mathrm{v}$ 
goes from~30\de\ to~60\de, 
and also about~3\% as $\theta_\mathrm{v}$ goes from~60\de\ to~90\de.  This
is because cocoons have geometries similar to prolate spheroids,
inside which the poloidal momentum flux is higher than the toroidal
one.  Cocoon \mfs\ are thus mainly stretched along the polar direction
(the jet axis) which projection onto the line of sight is proportional to
$\cos(\theta_\mathrm{v})$.

We see 
the main 
features of the $|\chi_\mathrm{B}|$ histograms are shaped
during the early expansion phase of sources, particularly for
$\theta_\mathrm{v} \ga\,$60\de. In this case, magnetic
fields tend to align with the jet axis as sources grow.  For 
$\theta_\mathrm{v} =\,$30\de, 
on the other hand, 
$\chi_\mathrm{B}$ is distributed nearly
isotropically. The fractional polarisation is broadly distributed
around about 30-40~per cent, and decreases in time.

\section*{Acknowledgements}

The software used in these investigations was in part developed by
the DOE-supported ASC / Alliance Center for Astrophysical Thermonuclear
Flashes at the University of Chicago.  The authors wish to thank
Dongwook~Lee for the 3D-USM-MHD solver of Flash3.1, and also to
Malcolm Longair, Robert Laing, Julia Riley and Eric Blackman for
useful discussions and comments that helped to improve this paper.
MHE acknowledges financial support from CONACyT (The Mexican National
Council of Science and Technology, 196898/217314).


\bsp

\label{lastpage}
 
\end{document}